\documentclass{elsart}

\usepackage{graphicx}
\usepackage{epsfig}

\usepackage{amssymb}
\usepackage{color}

\begin{document}
\begin{frontmatter}

\title{Cold Uniform Matter and Neutron Stars in the Quark-Meson-Coupling Model}

\author[1,2]{J. Rikovska Stone}
\address[1] {Department of Physics, University Oxford,\\
  Oxford OX1 3PU, United Kingdom}
  
  \address[2] {Department of Chemistry and Biochemistry,\\
  University of Maryland, College Park, MD 20742, USA}

\author[4]{P.A.M. Guichon\corauthref{cor1}}
\corauth[cor1]{corresponding author: pierre.guichon@cea.fr}

\address[4] {SPhN-DAPNIA, CEA Saclay, F91191 Gif sur Yvette, France}

\author[6,5] {H.H. Matevosyan}
 
\address[6] {Louisiana State University, Department of Physics \\ 
  \& Astronomy, 202 Nicholson Hall, Tower Dr., LA 70803, USA }
 
\author[5] {A.W. Thomas}

\address[5] {Thomas Jefferson National Accelerator Facility,\\ 
  12000 Jefferson Ave., Newport News, VA 23606, USA}

\begin{abstract}
A new density dependent effective baryon-baryon interaction has been 
recently derived from the quark-meson-coupling (QMC) model, offering 
impressive results in application to finite nuclei and dense baryon 
matter. This self-consistent, relativistic quark-level approach is used 
to construct the Equation of State (EoS) and to calculate key properties of 
high density matter and cold, slowly rotating neutron stars. 
The results include predictions for the maximum mass of neutron star models, 
together with the corresponding radius and central density, as well the 
properties of neutron stars with mass of order 1.4 $M_\odot$. 
The cooling mechanism allowed by the QMC EoS is explored and the parameters 
relevant to slow rotation, namely the moment of inertia and the period of 
rotation investigated. The results of the calculation, which are found to 
be in  
good agreement with available observational data, are compared with the 
predictions of more traditional EoS, based on the A18+$\delta$v+UIX$^*$ 
and modified Reid soft core potentials, the Skyrme SkM$^*$ interaction and 
a relativistic mean field (RMF) models for a hybrid stars including
quark matter. The QMC EoS provides cold neutron star models with 
maximum mass 1.9--2.1 M$_\odot$, with central density less than 6 times 
nuclear saturation density ($n_{0}= 0.16 {\rm\ fm}^{-3}$) and offers a 
consistent description of the stellar mass up to this density limit. In 
contrast with other models, QMC predicts no hyperon contribution at 
densities lower than  $3n_0$, for matter in $\beta$-equilibrium. At 
higher densities, $\Xi^{-,0}$ and $\Lambda$ hyperons are present. 
The absence of lighter $\Sigma^{\pm,0}$ hyperons is understood as a 
consequence of antisymmetrisation, together with the  implementation of 
the color hyperfine interaction in the response of the quark bag to the 
nuclear scalar field.         
\end{abstract}

\end{frontmatter}


%
\section{Introduction}
\label{sec:intro}

Nuclear forces play an important role in many stellar environments, 
acting in concert with gravitational forces 
to form compact objects. For example, the observed  
maximum mass of cold neutron stars cannot be 
explained without considering the pressure, arising from the strong 
nuclear force, which opposes gravitational collapse. 
It follows that strongly interacting matter is one of the fundamental 
systems that we have to study in order to 
understand stellar phenomena as well as the properties of finite nuclei. 

The properties of stellar matter are strongly density dependent. The 
theoretical framework for describing the properties of strongly interacting  
matter is the Equation of State (EoS), which relates the pressure 
and total energy density of the system. These quantities are derived from 
the density dependence of the 
energy per particle of the system, 
calculated using a model for the baryon-baryon interaction 
between particles present in the matter. 
There is a large variety of EoS available in the 
literature (see e.g. \cite{lat00,hei00,lat01,rik03}), 
based on very different assumptions concerning the 
nature of the interactions governing the stellar matter. 
Ways of constraining the available 
EoS are being actively sought 
(see e.g. \cite{rik03,lat05a,lat05b,pod05,kla06}) but no 
unambiguous findings have been reported as yet. 
The study of the properties of high density matter and of cold neutron 
stars provides an attractive laboratory for seeking such a constraint. 
However, the fundamental question 
arises as to the limit of applicability of any of the 
existing EoS at densities as high 
as $6\div10\,n_0$, required, for example, for calculating 
the maximum mass of neutron star models.  

At sufficiently high density, nuclear matter will most likely be 
composed not just of nucleons and leptons but also  
strange baryons and boson 
condensates~\cite{Weber:2004kj,Schaffner-Bielich:2004ch,Maruyama:2005yp,Tatsumi:1995is}. 
More importantly, it is 
unlikely that these particles will 
maintain their identity as confined entities of quarks, 
but the interaction between overlapping quark bags 
will result in partial or full deconfinement~\cite{Lawley:2006ps,Lawley:2005ru}. 
Therefore a realistic high density EoS should contain 
all this physics. Numerical extrapolation of a nucleon-only EoS to 
high densities, although it may appear to work (mechanically), 
is dangerous and may obscure some important effects. This is true even at lower 
densities, in the region around twice nuclear saturation density,
as some models predict low threshold 
densities for the creation of hyperons in beta equilibrium. The investigation of this effect 
on non-equilibrium matter is expected to be crucial for modelling  
core-collapse supernovae, as the density at bounce after the 
original collapse is around $2\div2.5\,n_0$ 
and the presence of hyperons will soften the EoS, thus assisting the explosion.

The relativistic formulation of the QMC model~\cite{gui04,gui06} 
offers a unique opportunity to 
self-consistently probe the composition of high density matter. 
In this work, the matter which is investigated is taken to be in 
$\beta$-equilibrium, including the full baryon octet as well as 
electrons and muons. 
The density and temperature dependent thresholds and production 
rates of hyperons in non-equilibrium 
stellar matter, relevant for core-collapse supernova physics, 
will be the subject of a forthcoming 
publication.

The general QMC model is briefly described in 
Section~\ref{sec:qmc}, followed by a 
description of its application to uniform nuclear matter 
at zero temperature and in 
$\beta$-equilibrium (Section~\ref{sec:nmatter}). 
The basic features of cold, slowly rotating 
neutron stars are given in Section~\ref{sec:nstars}. 
Details of QMC neutron star models and 
some comparison with observational data and other model calculations 
are given in Section~\ref{sec:resdis},   
with a summary of our main conclusions in Section~\ref{sec:concl}.        

\section{QMC model}
\label{sec:qmc}
\subsection{The physics of the model}
The formulation of the QMC model~\cite{gui88} which we use in this work has been
presented in a previous paper \cite{gui06}. There it was shown that it is
possible to build a nuclear Hamiltonian which is consistent with relativity
and which can be used at high density. The predictions of this
model, which represents a significant improvement over earlier 
work~\cite{gui04}, because the need to expand about $<\sigma> = 0$ was 
removed, have been sucessfully  
confronted with the phenomenology of
finite nuclei. In this section we rely on 
the results of Ref.~\cite{gui06} and
generalize the formalism to nuclear matter containing an 
arbitrary mixture of the octet members, $N,\Lambda,\Sigma,\Xi$.

We recall that, 
in the QMC model, the nuclear system
is represented as a collection 
of quark bags. These bags may contain
strange as well as non strange quarks, so the framework 
is well adapted
to describing any kind of baryon. The crucial hypothesis is that in
nuclear matter the baryons retain their individuality, or more precisely,
they remain the pertinent degrees of freedom (or quasi-particles) 
even in dense nuclear 
matter. From this point of view the usual statement
that the bags should not overlap is 
clearly too restrictive. It gives
too much weight to the description of the internal 
structure within the bag model, while the latter is in fact used only to 
infer the interactions 
between the quasi-particles. In this respect, we point out that the
bag model is an 
effective realisation of confinement, which must not
be taken too literally. The QCD 
lattice simulations of Bissey {\it et al.}~\cite{Bissey:2005sk}
indicate that the true confinement picture 
is closer to a Y-shaped color string
attached to the quarks. Outside this relatively 
thin string one has
the ordinary, non-perturbative vacuum, where the quarks from other 
hadrons can pass without disturbing the structure very much. So, 
even though the bag model 
imposes a strict condition, which prevents the quarks
from travelling through its boundary, 
this must be seen as the average
representation of a more complex situation and  one should
not attribute a deep physical meaning to the surface of the cavity 
nor to its size. In particular, estimating the density at which the 
QMC approximation breaks down as the reciprocal of the 
bag volume is certainly too 
pessimistic. Of course, there is
a density above which the QMC model becomes 
inadequate but in this
work we assume that this critical density is large enough that we 
can use the model to predict the properties of neutron stars. 

The salient feature of the QMC model is that the interactions are
generated by the exchange 
of mesons coupled locally to the quarks.
In a literal interpretation of the bag model, 
where only quarks and
gluons can live inside the cavity, this coupling would be unnatural.
Again, we must return to the more realistic underlying picture where the quarks
are attached to 
a string but otherwise move in the non-perturbative
QCD vacuum. There, nothing prevents 
them from feeling the vacuum fluctuations, which we describe by meson fields. 
As before we limit our considerations to the 
$\sigma,\,\omega$ and $\rho$ mesons. We 
also estimate, in a simplified manner, the effect of single pion exchange. 
At this point it is useful to recall that the $\sigma$ 
meson used here is not the chiral 
partner of the pion. It is a chiral invariant scalar field, 
which mocks up the correlated multi-pion 
exchanges in the scalar-isoscalar
channel.

\subsection{Effective mass and energy of 
a bag in the nuclear medium}

In Ref.~\cite{gui96} we used the Born-Oppenheimer (BO) approximation
to derive the energy 
of a quark bag coupled to the nuclear mean 
fields associated with the $\sigma,\,\omega$ and $\rho$ mesons. 
Given the position and velocity of a bag, we obtained its energy 
by solving the field equations for the confined quarks.
As the BO approximation requires that both position 
and velocity are known,
this energy is of course classical and canonical quantization is  
performed
after the full Hamiltonian has been built. 
As our previous work was limited to 
nuclear matter and finite nuclei built primarily of nucleons, 
we now generalize our results to
include hyperons. Here we limit 
our considerations to the spin
1/2 SU(3) octet $(N,\Lambda,\Sigma,\Xi)$ and therefore a baryon  
can be specified by $|f>=|tms>$, 
with $t,m$ the isospin and its projection 
and $s$ the strangeness -- see Table \ref{cap:QMC-Octet}. 
\begin{table}
\begin{center}
\begin{tabular}{|c|c|c|c|c|c|c|c|c|}
\hline 
&
$p$&
$n$&
$\Lambda$&
$\Sigma^{-}$&
$\Sigma^{0}$&

$\Sigma^{+}$&
$\Xi^{-}$&
$\Xi^{0}$\tabularnewline
\hline
\hline 
$t$&
$\frac{1}{2}$&

$\frac{1}{2}$&
0&
1&
1&
1&
$\frac{1}{2}$&
$\frac{1}{2}$\tabularnewline
\hline 
$m$&

$\frac{1}{2}$&
$-\frac{1}{2}$&
0&
-1&
0&
1&
$-\frac{1}{2}$&
$\frac{1}{2}$\tabularnewline
\hline 
$s$&
0&
0&
-1&
-1&
-1&
-1&
-2&
-2\tabularnewline
\hline
\end{tabular}
\end{center}
\caption{\label{cap:QMC-Octet}Quantum numbers of the octet members.}
\end{table}
Our working hypothesis is that the strange quark is not coupled to
the meson fields, 
which can be seen as a consequence of the fact
that the mesons represent correlated pion exchanges. 
This is also
the natural explanation for the small spin-orbit splitting   
observed in $\Lambda$ 
hypernuclei, given that all the $\Lambda$ spin is carried
by the strange quark. We set the $u$ and $d$ 
quark masses to zero and 
assume that their coupling to the meson fields respects isospin 
symmetry. Then the energy of a baryon of flavour $f$ with position
$\vec{R}$ and momentum $\vec{P}$ 
has the form \footnote{We use the system of units such that $\hslash =c=1$.}
\begin{equation}
E=\sqrt{P^{2}+M_{f}(\sigma)^{2}}+g_{\omega}^{f}\omega+
g_{\rho}\vec{b}.\vec{I}^{t}
\label{eq:QMC1}
\end{equation}
where $M_{f}(\sigma)$ is the effective mass, that is the 
energy of the corresponding quark bag in its rest frame, in  
the $\sigma$ field, and $\vec{I}^{t}$ is
the isospin operator for isospin $t$, 
defined as 
\begin{equation}
<tms|\sum_{q}\frac{\vec{\tau_{q}}}{2}|t'm's'>=
\delta(tt')\delta(ss')\vec{I}_{mm'}^{t}
\label{eq:QMC2}
\end{equation}
with $\vec{\tau}_q$ the Pauli matrix acting on the $u$ and $d$ quarks.
Apart from the $\rho$ coupling, for which we denote the field $b_{\alpha}$,  
with $\alpha=(-1,0,1)$ the isospin index, 
the energy is diagonal in flavour space. 
Note that the flavour dependence of the $\rho$ coupling is
entirely contained in $\vec{I}^{t}$. 
In Eq.~(\ref{eq:QMC1}) the meson
fields are evaluated at $\vec{r}=\vec{R}$, the center 
of the bag%
\footnote{The variation of the field over the confinement cavity generates the
spin-orbit interaction, as derived in Ref.~\cite{gui06}.%
} and, in comparison with Ref.~\cite{gui06}, we have dropped the spin-orbit 
interaction -- 
since in the following we shall consider only the case of uniform
matter. 
The flavour dependence of the $\omega$ coupling is fixed by the number
of non-strange quarks 
in the baryon, that is:
\begin{equation}
g_{\omega}^{f}=
w_{f}^{\omega}g_{\omega}=(1+\frac{s}{3})g_{\omega}\label{eq:QMC3}
\end{equation}
with $g_{\omega}$ the $\omega-N$ coupling constant. As shown in the 
Appendix, the effective mass is well approximated by a 
quadratic expansion
\begin{equation}
M_{f}(\sigma)=M_{f}-w_{f}^{\sigma}g_{\sigma}\sigma+
\frac{d}{2}\tilde{w}_{f}^{\sigma}\left(g_{\sigma}\sigma\right)^{2}
\label{eq:QMC4}
\end{equation}
where $g_{\sigma}$ is the free $\sigma-N$ coupling constant. The
scalar polarisability, $d$, describes the response of the nucleon to
the applied scalar field. It is at the origin of the many-body forces in the
QMC model. The corresponding scalar polarisability for baryon $f$ is $d\tilde{w}_f^\sigma (w_f^\sigma)^{-2}$. The weights $w_{f}^{\sigma},\tilde{w}_{f}^{\sigma}$ control
the flavour dependence and in first approximation 
$w_{f}^{\sigma}\simeq\tilde{w}_{f}^{\sigma}\simeq1+s/3.$
The hyperfine color interaction breaks this relation and the
exact values, which depend on the free nucleon radius, $R_{N}^{free}$,
are given in the Appendix.

\subsection{Hamiltonian}

The total energy of the nuclear system is the sum of the baryon energies
(\ref{eq:QMC1}) and of the energy stored in the meson fields:
\begin{eqnarray}
E_{tot} & = & \sum_{i=1,A}E(i)+E_{mesons}\nonumber \\
E_{mesons} & = & \frac{1}{2}\int d\vec{r}\left[\left(\nabla\sigma\right)^{2}+
m_{\sigma}^{2}\sigma^{2}\right]\nonumber \\
 &  & -\frac{1}{2}\int d\vec{r}\left[\left(\nabla\omega\right)^{2}+
m_{\omega}^{2}\omega^{2}\right]-\frac{1}{2}\int d\vec{r}
\left[\left(\nabla\vec{b}\right)^{2}+m_{\rho}^{2}\vec{b}^{2}\right]
\label{eq:QMC5}
\end{eqnarray}
with $m_{\sigma},m_{\omega},m_{\rho}$ the meson masses. As usual,  
we consider only the time component of the vector fields.
By hypothesis, the meson fields are time independent. Therefore the
classical Hamiltonian of the nuclear system is simply
\begin{equation}
H(R_{i},P_{i})=E_{tot}(R_{i},P_{i},\sigma\to\sigma_{sol},\omega\to\omega_{sol},\vec{b}\to\vec{b}_{sol})
\label{eq:QMC6}
\end{equation}
where $\sigma_{sol},\omega_{sol},\vec{b}_{sol}$ are the the solutions 
of the meson equations of motion:
\begin{equation}
\frac{\delta E}{\delta\sigma}=\frac{\delta E}{\delta\omega}=
\frac{\delta E}{\delta b_{\alpha}}=0.\label{eq:QMC7}
\end{equation}
The equations for the $\omega,\vec{b}$ fields, which are linear,
present no difficulty. By contrast, the $\sigma$ field equation is
highly non-linear, because of the scalar polarisability term in the effective
mass. In Ref.~\cite{gui06}, we proposed to solve it 
approximately by writing:
\begin{equation}
\sigma=\bar{\sigma}+\delta\sigma,\ \ \ \bar{\sigma}=<\sigma> \label{eq:QMC8}
\end{equation}
where, $<\cdots >$,  
denotes the nuclear ground state expectation value, and
we considered the deviation, $\delta\sigma$, as a small quantity. We
refer to Ref.~\cite{gui06} for the details of the derivation and 
here we simply recall
the results. The piece of the Hamiltonian which depends on 
$\sigma$ may be written:
\begin{equation}
H_{\sigma}=\int d\vec{r}\,\left[K(\bar{\sigma})-\frac{1}{2}
\bar{\sigma}<\frac{\partial K}{\partial\bar\sigma}>+
\frac{1}{2}\delta\sigma\left(\frac{\partial K}{\partial\bar\sigma}-
<\frac{\partial K}{\partial\bar\sigma}>\right)\right]
\label{eq:QMC9}
\end{equation}
where normal ordering is implicit. 
The one body kinetic energy operator, $K(\bar{\sigma})$, is defined as ($V$
is the normalization volume):
\begin{equation}
K(\bar{\sigma})=\frac{1}{2V}\sum_{\vec{k},\vec{k}',f}
e^{i(\vec{k}-\vec{k}').\vec{r}}\left(\sqrt{k^{2}+
M_{f}[\bar{\sigma}(\vec{r})]^{2}}+\sqrt{k'^{2}+
M_{f}[\bar{\sigma}(\vec{r})]^{2}}\right)a_{\vec{k}f}^{\dagger}a_{\vec{k}'f}
\label{eq:QMC10}
\end{equation}
with $a_{\vec{k}f}$ the destruction operator for a baryon of momentum 
$\vec{k}$ and flavour $f$, the spin label being omitted for simplicity.
Note that the mean field approximation amounts to setting $\delta\sigma=0$
in the Hamiltonian (\ref{eq:QMC9}).

The meson field solution in the general case is of little interest
to us, so we directly give the solution for a uniform system in which
case $\bar{\sigma}$ is a constant determined by the self consistent
equation:
\begin{equation}
\bar{\sigma}=-\frac{1}{m_{\sigma}^{2}}<\frac{\partial K}{\partial\bar\sigma}>
\label{eq:QMC11}
\end{equation}
which must be solved numerically and the fluctuation $\delta\sigma$
is given by:
\begin{equation}
\delta\sigma(\vec{r})=\int d\vec{r}'\frac{d\vec{q}}{(2\pi)^{3}}
\frac{e^{i\vec{q}.(\vec{r}-\vec{r}')}}{q^{2}+
\tilde{m}_{\sigma}^{2}}\left(-\frac{\partial K}{\partial\bar\sigma}(\vec{r}')+
<\frac{\partial K}{\partial\bar\sigma}>\right)
\label{eq:QMC12}
\end{equation}
where the effective $\sigma$ mass is:
\begin{equation}
\tilde{m}_{\sigma}^{2}=m_{\sigma}^{2}+
<\frac{\partial^{2}K}{\partial\bar{\sigma}^{2}}>.
\label{eq:QMC15}
\end{equation}

\subsection{Energy density in the Hartree Fock approximation}

The last step in our formal development is to use 
Eqs.~(\ref{eq:QMC9}-\ref{eq:QMC12})
to compute the energy density of nuclear matter in the Hartree-Fock
approximation. The ground state of the system is specified by a set
of Fermi levels, $\{ k_{F}(f),f=p,n...\}$, and their corresponding
baryonic densities $\{ n_{f}\}$ with $n_B=\sum_{f}n_{f}$. One has:
\begin{equation}
<K(\bar{\sigma})>=\frac{2}{(2\pi)^{3}}\sum_{f}\int_{o}^{k_{F}(f)}
d\vec{k}\sqrt{k^{2}+M_{f}^{2}(\bar{\sigma})}
\label{eq:QMC13}
\end{equation}
and the energy density takes the form  
\begin{eqnarray}
\frac{<H_{\sigma}>}{V} & = & <K(\bar{\sigma})>+\frac{1}{2m_{\sigma}^{2}}
\left(<\frac{\partial K}{\partial\bar\sigma}>\right)^{2}+
\frac{1}{(2\pi)^{6}}\sum_{f}\int_{0}^{k_{F}(f)}d\vec{k}_{1}d\vec{k}_{2}
\nonumber \\
 &  & \frac{1}{(\vec{k}_{1}-\vec{k}_{2})^{2}+\tilde{m}_{\sigma}^{2}}
\frac{\partial}{\partial\bar{\sigma}}\sqrt{k_{1}^{2}+
M^{2}(\bar{\sigma})}\frac{\partial}{\partial\bar{\sigma}}
\sqrt{k_{2}^{2}+M^{2}(\bar{\sigma})}.
\label{eq:QMC14}
\end{eqnarray}
Finally we must add the contributions $<V_{\omega}> and <V_{\rho}>$, 
corresponding
to $\omega$ and $\rho$ exchange. The derivation follows the
same line as for the $\sigma$ exchange but is much simpler and gives
the exact solution because these interactions are purely 2-body .
One gets:
\begin{eqnarray}
&&\frac{<V_{\omega}>}{V} =  \frac{G_{\omega}}{2}\left(\sum_{f}w_{f}^{\omega}n_{f}\right)^{2}
\nonumber \\
&& - G_{\omega}\sum_{f}\left(w_{f}^{\omega}\right)^{2}\frac{1}{(2\pi)^{6}}
\int_{0}^{k_{F}(f)}d\vec{k}_{1}d\vec{k}_{2}
\frac{m_{\omega}^{2}}{(\vec{k}_{1}-\vec{k}_{2})^{2}+m_{\omega}^{2}}
\label{eq:QMC16}\\
&&\frac{<V_{\rho}>}{V}  =  \frac{G_{\rho}}{2}\left(\sum_{tms}mn_{tms}\right)^{2}\nonumber \\
&& -  G_{\rho}\sum_{tmm's}\vec{I}_{mm'}^{t}.\vec{I}_{m'm}^{t}\frac{1}{(2\pi)^{6}}
\int_{0}^{k_{F}(tms)}d\vec{k}_{1}\int_{0}^{k_{F}(tm's)}d\vec{k}_{2}\frac{m_{\rho}^{2}}{(\vec{k}_{1}-\vec{k}_{2})^{2}+m_{\rho}^{2}}
\label{eq:QMC17}
\end{eqnarray}
As usual we define:
\begin{equation}
G_{\sigma}=\frac{g_{\sigma}^{2}}{m_{\sigma}^{2}},\, G_{\omega}=
\frac{g_{\omega}^{2}}{m_{\omega}^{2}},\, G_{\rho}=
\frac{g_{\rho}^{2}}{m_{\rho}^{2}}.
\label{eq:QMC18}
\end{equation}

The $\sigma,\omega$ and $\rho$ mesons represent the multi-pion exchanges
that we believe to be the most relevant ones. However, this cannot
take into account the long ranged single pion exchange. In the
Hartree Fock approximation, the latter contributes only through the exchange
(Fock) term, so its contribution to the energy is modest. Nevertheless,  
we think it is worthwhile to study its impact on our results because, 
as a consequence of its long range, the pion contribution 
to the energy has, as compared to the heavy mesons, 
a specific  density dependence. With this in mind, 
we generalize the  expression of Ref.~\cite{Krein:1998vc} so as 
to include the contribution of the full octet. We get:
\begin{eqnarray}
\frac{<V_{\pi}>}{V} & = & \frac{1}{n_B}\left(\frac{g_{A}}{2f_{\pi}}\right)^{2}
\left\{ J_{pp}+4J_{pn}+J_{nn}-\frac{24}{25}\left(J_{\Lambda,\Sigma^{-}}+
J_{\Lambda,\Sigma^{0}}+J_{\Lambda,\Sigma^{+}}\right)\right.
\nonumber \\
& + & \left.\frac{16}{25}\left(J_{\Sigma^{-}\Sigma^{-}}+
2J_{\Sigma^{-}\Sigma^{0}}+2J_{\Sigma^{+}\Sigma^{0}}+J_{\Sigma^{+}\Sigma^{+}}
\right)\right.
\nonumber \\
 & + & \left.\frac{1}{25}\left(J_{\Xi^{-}\Xi^{-}}+
4J_{\Xi^{-}\Xi^{0}}+J_{\Xi^{0}\Xi^{0}}\right)\right\} 
\label{eq:QMC19}
\end{eqnarray}
with :
\begin{eqnarray}
J_{ff'} & = & \frac{1}{(2\pi)^{6}}\int^{k_{F}(f)}\int^{k_{F}(f')}d\vec{p}
d\vec{p}'\left[1-\frac{m_{\pi}^{2}}{(\vec{p}-\vec{p}')^{2}+
m_{\pi}^{2}}\right].
\label{eq:QMC20}
\end{eqnarray}
In Eq.~(\ref{eq:QMC19}), $g_{A}=1.26$ is the axial coupling constant
of the nucleon, $f_{\pi}=93{\rm\ MeV}$ is the pion decay constant 
and $m_{\pi}$
its mass. In the loop integral $J_{ff'}$, the first term is the 
so-called contact term. In practice it can hardly be distinguished from
the short-ranged contributions of the heavy mesons and we omit it. 
We have checked that it 
would only induce a small readjustment
of $G_{\sigma},G_{\omega},G_{\rho}$. In the same spirit we have omitted
any $\pi$-baryon form factor in $J_{ff'}$, 
because the  effect of a form factor is
to cut off the large momenta of the loop and it therefore amounts
to a short ranged contribution. In summary, we keep only the long ranged Yukawa
piece of the pion exchange.

\subsection{Fixing the parameters}

The parameters of the model are the couplings $G_{\sigma},G_{\omega},G_{\rho}$,
the meson masses and the free nucleon radius. We set the $\pi,\omega,\rho$
masses to their physical values. We have tried several values of the
free nucleon radius, $R_{N}^{free}$, and found that this has very little
influence on our results. We therefore take $R_{N}^{free}=0.8{\rm\ fm}$, which
is a realistic value~\cite{Thomas:1982kv}. The $\sigma$ meson mass
is not well known because the width of the $\pi\pi$ resonance is
very large in the physical region. 
In our study of finite nuclei~\cite{gui06}, we found 
that $m_{\sigma}\sim700{\rm\ MeV}$
produced the best results, so this will be our prefered value. However,  
the sensitivity of the calculations of finite nuclei to the $\sigma$
meson mass comes from the fact that it controls the shape of the nuclear
surface, a factor which is irrelevant in the case of neutron stars.
Moreover, since the calculations for  finite nuclei are based 
on a non-relativistic,
low density approximation%
\footnote{By this we mean a density about or smaller 
than the density of ordinary
nuclear matter.%
} to the Hamiltonian (\ref{eq:QMC9}), we must be conservative about
the $\sigma$ meson mass when we use it in the high density
region. Therefore, to quantify the influence of this mass, we 
shall show some results for $m_{\sigma}=600{\rm\ MeV}.$ 

The remaining free parameters, $G_{\sigma},G_{\omega}$ and $G_{\rho}$ are
adjusted to reproduce the binding energy and the asymmetry energy of ordinary
nuclear matter at the saturation point. Since we want to sample the
stiffness of our equation of state, we allow some freedom in the way
the couplings are fixed. This is summarized in Table~\ref{cap:Couplings}.
First we set the pion contribution to zero and adjust the couplings
to reproduce the experimental values of the binding energy,  
$({\mathcal E}=-15.865{\rm\ MeV})$,  
and asymmetry energy, $(a_s=30{\rm\ MeV})$, of nuclear matter at the saturation
point ($n_0=0.16{\rm\ fm}^{-3}$). This defines the models QMC600 and QMC700, 
according to the value of the $\sigma$ meson mass. For all the other
models (QMC$\pi i,\, i=1,4$) we set $m_{\sigma}=700{\rm\ MeV}$ and the
effect of the pion is included. As can be seen in the last column
of Table~\ref{cap:Couplings}, the incompressibility modulus 
of both  QMC600 and QMC700 is above
the accepted range, $200\div300{\rm\ MeV}$. The most efficient way to decrease
$K_{\infty}$ is to add an attractive interaction which has a weak
density dependence at the saturation point. The pion Fock term, 
which is attractive and behaves
roughly as $\rho^{1/6}$ , is thus a natural candidate. This is
confirmed by the model QMC$\pi1$, where the pion contribution is
included and where $K_{\infty}$ drops from $340{\rm\ MeV}$ to $322{\rm\ MeV}$.
A trivial way to mock up the effect of an almost constant attractive
interaction is simply to decrease the absolute value of the experimental
binding energy. By setting ${\mathcal E}=-14.5{\rm\ MeV}$, a rather moderate
distortion of the experimental number, 
we already find $K_{\infty}=301{\rm\ MeV}.$
This defines QMC$\pi2.$ In order to sample a wider range of values of the
stiffness, we allow ourself to artificially vary both the pion
contribution and the binding energy. This leads to model QMC$\pi3$
(resp. QMC$\pi4$), where the pion contribution is multiplied by
1.5 (resp. 2) and the binding energy is set to 
$-14{\rm\ MeV}$ (resp $-13{\rm\ MeV}$).
This gives $K_{\infty}=283{\rm\ MeV}$ and $K_{\infty}=256{\rm\ MeV}$, 
respectively.
Note that increasing the pion contribution by a factor of 2 is not so
arbitrary, because we know that it is amplified by RPA correlations
\cite{chan06}. On the other hand, changing the experimental binding energy
is more difficult to justify and we take it only as a crude way to
sample the stiffness of the model, given its eventual 
application in the high density
region, far away from the point where ${\mathcal E}$ is measured. 
\begin{table}
\begin{center}\begin{tabular}{|c|c|c|c|c|c|c|c|}
\hline 
Model&
$m_{\sigma}({\rm\ MeV})$&
$\pi$ &
${\mathcal E}({\rm\ MeV})$&
$G_{\sigma}({\rm\ fm}^{2})$&
$G_{\omega}({\rm\ fm}^{2})$&
$G_{\rho}({\rm\ fm}^{2})$&
$K_{\infty}({\rm\ MeV})$\tabularnewline
\hline
\hline 
QMC600&
600&
0&
-15.865&
11.23&
7.31&
4.81&
344\tabularnewline
\hline 
QMC700&
700&
0&
-15.865&
11.33&
7.27&
4.56&
340\tabularnewline
\hline 
QMC$\pi1$&
700&
1&
-15.865&
10.64&
7.11&
3.96&
322\tabularnewline
\hline 
QMC$\pi2$&
700&
1&
-14.5&
10.22&
6.91&
3.90&
301\tabularnewline
\hline 
QMC$\pi3$&
700&
1.5&
-14&
9.69&
6.73&
3.57&
283\tabularnewline
\hline 
QMC$\pi4$&
700&
2&
-13&
8.97&
6.43&
3.22&
256\tabularnewline
\hline
\end{tabular}
\end{center}
\caption{\label{cap:Couplings}The couplings for different versions of the
model. The column $\pi$ is the number by which the pion contribution
has been multiplied. ${\mathcal E}$ is the binding energy of symmetric nuclear
matter and $K_{\infty}$ its incompressibility modulus. }
\end{table}

\section{Cold Uniform Matter}
\label{sec:nmatter}

In this section we determine the equation of state of uniform matter, 
which is directly relevant
for studies of the interior of neutron stars. 
The structure of the neutron stars is discussed in the next section.
Uniform matter in cold neutron stars is in a generalized beta-equilibrium, achieved over an extended period of time. All baryons of the octet
can be populated by succcessive weak interactions, regardless of their
strangeness~\cite{gle00}. We consider matter formed
by baryons of the octet, electrons and negative muons, with respective
densities $\{ n_{f},f=p,n,...\}$ and $\{ n_{e},n_{\mu}\}.$

The equilibrium
state minimises the total energy density, $\epsilon$, under the constraint
of baryon number conservation and electric neutrality. We write:
\begin{equation}
\epsilon=\epsilon_{B}(n_{p},...)+\epsilon_{e}(n_{e})+\epsilon_{\mu}(n_{\mu})
\label{eq:beta1}
\end{equation}
where the baryonic contribution
\begin{equation}
\epsilon_{B}(n_{p},...)=\frac{<H_{\sigma}+V_{\omega}+V_{\rho}+V_{\pi}>}{V}
\label{eq:beta2}
\end{equation}
is calculated from Eqs.~(\ref{eq:QMC14}-\ref{eq:QMC19}). 
It is related to the binding
energy per baryon, ${\mathcal E}$, by:
\begin{equation}
\epsilon_{B}(n_{p},...)=\sum_{f}({\mathcal E}+M_{f})n_{f}.
\label{eq:beta3}
\end{equation}
The baryonic pressure, ($P_{B}$), the incompressibility modulus, ($K_{\infty}$),
and the sound velocity, ($v_{s}$), of baryonic matter have the following
expressions:
\begin{equation}
P_{B}=n_{B}^{2}\frac{\partial}{\partial n_{B}}
\frac{\epsilon_{B}}{n_{B}},\,\,\, K_{\infty}=
9\frac{\partial P_{B}}{\partial n_{B}},\,\,\, 
v_{s}=\sqrt{\frac{n_{B}K_{\infty}}{9(\epsilon_{B}+P_{B})}}
\label{eq:beta4}
\end{equation}
where $n_{B}=\sum_{f}n_{f}$ and the derivative, with respect to $n_B$, 
is taken at constant fractions $(n_f/n_B,f=1,8)$. Similar expressions 
hold for each lepton.
In proton-neutron matter another important variable is the symmetry
energy, ${\mathcal S}$, defined as the difference between pure neutron
and symmetric matter :
\begin{equation}
{\mathcal S}(n_{B})={\mathcal E}(\left(n_{p}=0,\, n_{n}=
n_{B}\right)-{\mathcal E}\left(n_{p}=n_{B}/2,\, n_{n}=n_{B}/2\right).
\label{eq:beta5}
\end{equation}

For the energy density of the lepton $l$, of mass $m_{l}$ and density
$n_{l}$, we use the free Fermi gas expression:
\begin{equation}
\epsilon_{l}(n_{l})=\frac{2}{(2\pi)^{3}}\int^{k_{F}(l)}d\vec{k}
\sqrt{k^{2}+m_{l}^{2}},\,\,\,{\rm with}\, n_{l}=
\frac{k_{F}^{3}(l)}{3\pi^{2}}.
\label{eq:beta6}
\end{equation}

The equilibrium condition for a neutral system with baryon density
$n_{B}$ is 
\begin{eqnarray}
&&\delta [\epsilon_{B}(n_{p},...)+\epsilon_{e}(n_{e})+\epsilon_{\mu}(n_{\mu})
\nonumber\\
&& +\lambda(\sum_{f}n_{f}-n_{B})+\nu(\sum_{f}n_{f}q_{f}-(n_{e}+
n_{\mu}) ]=0
\label{eq:beta7}
\end{eqnarray}
where $q_{f}$ is the charge of the flavor $f$. The constraints are
implemented through the Lagrange multipliers $(\lambda,\nu)$ 
so the variation in Eq.~(\ref{eq:beta7})
amounts to independent variations of the densities, together with 
variations of $\lambda$ and $\nu$. If one defines the chemical potentials
as 
\begin{equation}
\mu_{f}=\frac{\partial\epsilon_{B}}{\partial n_{f}},\,\,\,\mu_{l}=
\frac{\partial\epsilon_{l}}{\partial n_{l}}=\sqrt{k_{F}^{2}(l)+m_{l}^{2}}
\label{eq:beta8}
\end{equation}
the equilibrium equations become:
\begin{eqnarray}
\mu_{i}+\lambda+\nu q_{i} & = & 0,\,\,\, f=p,...,\label{eq:beta9}\\
\mu_{e}-\nu & = & 0,\label{eq:beta10}\\
\mu_{\mu}-\nu & = & 0,\label{eq:beta11}\\
\sum_{f}n_{f}-n_{B} & = & 0,\label{eq:beta12}\\
\sum_{f}n_{f}q_{f}-(n_{e}+n_{\mu}) & = & 0.\label{eq:beta13}
\end{eqnarray}
This is a system of non-linear equations for 
$\{ n_{p},...,n_{e},n_{\mu},\lambda,\nu\}.$
It is usual to eliminate the Lagrange multipliers, 
$(\lambda,\nu)$, using Eqs.~(\ref{eq:beta10})
and (\ref{eq:beta9}), with $f=neutron.$ However, for a given
value of $n_{B}$, the equilibrium equation, Eq.~(\ref{eq:beta7}), generally
implies that some of the densities vanish and therefore that the equations
generated by their variation drop from the 
system (\ref{eq:beta9}-\ref{eq:beta11}) --  
because there is nothing to vary! In particular, substituting $\nu=\mu_{e}$
in Eqs.~(\ref{eq:beta9}) is not valid when the electron disappears
from the system. The equations obtained by this substitution may have
no solution in the deleptonized region, since one has forced
$\nu=0$. To correct for this, one must restore $\nu$ as an independent
variable when one reaches this region. This is technically inconvenient
and we found it is much simpler to blindly solve 
the full system of equations, (\ref{eq:beta9}-\ref{eq:beta13}),  
for the set $\{ n_{p},...,n_{e},n_{\mu},\lambda,\nu\}$. The only
simplification which is not dangerous is the elimination of the muon
density in favor of the electron density by combining 
Eqs.~(\ref{eq:beta10},\ref{eq:beta11})
to write $\mu_{\mu}=\mu_{e}$, which is solved by 
\begin{equation}
k_{F}(\mu)=\Re\sqrt{k_{F}(e)^{2}+m_{e}^{2}-m_{\mu}^{2}}
\label{eq:beta14}
\end{equation}
where $\Re$ denotes the real part. This is always correct because
if the electron density vanishes then so does the muon density and
the relation (\ref{eq:beta14}) reduces to $0=0.$ 

To solve the system (\ref{eq:beta9}-\ref{eq:beta13}), let us define
the set of relative concentrations (note that the lepton concentrations
are also defined with respect to $n_{B}$)
\begin{equation}
Y=\{ y_{i}\}=\left\{ \frac{n_{p}}{n_{B}},\frac{n_{n}}{n_{B}},
\frac{n_{\Lambda}}{n_{B}},...,\frac{n_{e}}{n_{B}},
\frac{n_{\mu}}{n_{B}}\right\} .
\label{eq:beta15}
\end{equation}
Each member, $y_{i}$, of the set $Y$ is associated with an equation
$E_{i}$ , the one among Eqs.~(\ref{eq:beta9},\ref{eq:beta11}) which
is obtained by taking the variation of $\epsilon$ with respect to
$\rho_{i}=y_{i}n_{B}$. Let us assume that a solution, $Y_{0}$, has
been found at some  baryon density, $n_{B}=n_{0}$. One first tests
whether the threshold for some species, $i$, is crossed when $n_{0}$
is incremented by $\delta n$. Since $y_{i}=0$ below and across the
theshold, the condition for the appearance of the 
species is that $E_{i}(n_{B},Y_{0})$
changes its sign between $n_{0}$ and $n_{0}+\delta n$. If this happens,  
the equation $E_{i}$ is added to the system. If the threshold condition
is met for several species, all the corresponding equations are added.
The system is then solved numerically at $n_{B}=n_{0}+\delta n_{0}$,  
using $Y_{0}$ as a first approximation. When a concentration drops
below a given small number, $\eta$, it is set to zero and the corresponding
equation is removed from the system. The value of $\eta$ depends
on the accuracy of the solution and in our calculation it has been
set to $10^{-4}$. We have checked that $\eta=10^{-3}$ gives the
same result. We solve the system with the initial condition that at $n_{B}=0$
the matter contains only neutrons. Once the equilibrium solution $Y(n_{B})$
has been found for the desired range of baryon density, typically
$n_{B}=0\div1.2{\rm fm }^{-3}$, it is used to compute the equilibrium
total energy density, $\epsilon(n_{B})=\epsilon(y_{p}n_{B},...,y_{\mu}n_{B})$. The corresponding total pressure, $P(n_{B})$, is computed as 
the sum of the baryon pressure, (\ref{eq:beta4}), and of the lepton pressures. 
Using the equilibrium equations, Eqs.~(\ref{eq:beta9}-\ref{eq:beta13}), 
it is straightforward to show that the total pressure can also be 
evaluated as (note the total derivative):
\begin{equation}
P(n_{B})=n_{B}^{2}\frac{d}{dn_{B}}\frac{\epsilon(n_{B})}{n_{B}}.
\label{eq:beta16}
\end{equation} 
As a numerical check, we used both expressions, which did indeed yield 
the same result within the numerical accuracy.

\section{Neutron Stars}
\label{sec:nstars}
A neutron star is an object composed of matter at densities 
ranging from
that of terrestrial iron up to several times that of nuclear matter. 
In cold neutron stars, as the baryon  density $n_{\rm B}$ increases from 
about 0.75 $n_{\rm 0}$ up to $2\div 3 n_{\rm 0}$, 
 stellar matter 
becomes a homogeneous system of
unbound neutrons, protons, electrons and muons and, if enough time is
allowed, will develop $\beta$-equilibrium with respect to the 
weak interactions. All components that are present on a
timescale longer than the life-time of the system take part in
equilibrium. For example, neutrinos created in weak processes in a cold
neutron star do not contribute to the equilibrium conditions as they escape
rapidly. At even higher densities, heavier
mesons and strange baryons play a role (see e.g. Ref.~\cite{arn77}
and references therein,~\cite{pan71,bal99,wir93,hof01}). 
Ultimately, at the center
of the star, a quark matter phase may appear, either alone or coexisting with
hadronic matter~ \cite{gle00,iid98,bur02,men03}. 
Below $n_{\rm B}$ $\sim$ 0.75 $n_{\rm 0}$, the matter forms the inner 
and outer crust of the star, with inhomogeneities consisting of nucleons 
arranged on a lattice, as well as neutron and electron gases.  
In this density region the QMC Equation of State (EoS) needs to be 
matched onto other equations of state, reflecting the composition of
matter at those densities. The Baym-Bethe-Pethick (BBP)
\cite{bay71a} and Baym-Pethick-Sutherland (BPS) EoS~\cite{bay71b} 
are used in this work.

Setting up an EoS over the full density range, up to at least 6 $n_{\rm 0}$, 
allows calculation of one of the most important
observables of neutron stars, the maximum gravitational mass, 
$M_{\rm g}^{\rm max}$, and the corresponding radius, $R_{\rm max}$. The
most accurately measured masses of neutron stars were, until very recently,
consistent with the range 1.26 to 1.45 M$_\odot$~\cite{lat05a}. 
However, Nice
et al.~\cite{nic05} recently provided a dramatic result for the 
gravitational mass of the
PSR J0751+1807 millisecond pulsar (in a binary system with a helium white dwarf), M$_{\rm g}$~=~2.1$\pm $ 0.2 M$_\odot$ (with 68\% confidence) or 2.1$^{\rm +0.4}_{\rm -0.5}$ with 95\% confidence  
which makes
it the most massive pulsar measured. This observation potentially offers one of
the most stringent tests for the EoS used in the calculation of 
cold neutron stars. 
It
also sets an upper limit to the mass density, or equivalently, the energy
density, inside the star~\cite{lat05a}. A lower limit to the mass density can
be derived using the latest data on the largest observed redshift from a
neutron star, combined with its observational gravitational mass.

The calculated maximum mass is 
determined to large extent by the high density EoS. It has been argued 
that extrapolation of the nucleon-based EOS, built using an effective 
force (e.g. non-relativistic Skyrme~\cite{rik03}, relativistic 
mean field~\cite{gle00}) or phenomenological interactions 
(such as A18+$\delta v$+
UIX$^*$ (APR)~\cite{akm98}) to densities 
corresponding to $M_{\rm g}^{\rm max}$, may not be unreasonable~\cite{akm98,cha97} 
and that the error made by
not including the heavy baryons and possible quarks in the calculation may not
be significant. On the other hand, we note that in the hybrid calculations of 
Lawley {\it et al.}~\cite{Lawley:2006ps}, the maximum mass of a neutron star 
was significantly reduced. In any case, we will explore 
the consequences of this extrapolation 
by using the QMC EoS and taking into account the different composition of 
the homogeneous matter, starting from a nucleon-only model and then 
including the full baryon octet. We shall also comment shortly on 
the presence  of exotics, such as penta-quarks and 6-quarks bags, 
in high density matter.  Furthermore, neutron star models at around 
the `canonical' mass of $1.4 M_\odot$, with central densities of 
the order $2\div2.5\,n_0$  
and lower, are also studied as representative cases for nucleon-only based EoS.

\subsection{Cold non-rotational neutron star models}
\label{sec:coldnonrot}
The gravitational mass and the radius are calculated using a 
tabulated form of the composite EoS with a chosen QMC interaction. 
The Tolman-Oppenheimer-Volkov 
equation 
\cite{tol34,opp39}
\begin{equation}
  \frac{dP}{dr} 
  = 
  -\frac{Gm\rho}{r^2} \
  \frac{\left(1 + {P}/{\rho} \right) 
  \left(1 + {4\pi r^3P}/{m}\right)}{1-{2Gm}/{r}}
\label{eq:TOV}
\end{equation}
is integrated with
\begin{equation}
  m(r) 
  = 
  \int_0^r 4\pi {r^\prime}^2\rho (r^\prime)\,dr^\prime,
\label{eq:relmas}
\end{equation}
where $\rho$ is the total mass density and $G$ is the gravitational constant, 
in order to obtain sequences of neutron-star models corresponding to  
any specified central mass density. The solution gives 
directly the radius, $R$, of
the star (the surface being at the location where the pressure vanishes) 
and the corresponding value for the total gravitational mass $M_g=m(R)$.

It is also important to calculate some other properties of these
neutron star models. The total baryon number $A$ is given by
\begin{equation}
  A
  =
  \int_0^R \frac{4 \pi r^2 n_{\rm B}(r) dr}
  {\left( 1 - {2Gm(r)}/{r }\right)^{1/2}} \quad .
\label{eq:totbarnum}
\end{equation}
The total baryon number, $A$, multiplied by the atomic mass unit, 
931.50 {\rm\ MeV}, defines the \textit{baryonic mass} M${\rm _0}$. 
The binding energy released in a supernova core-collapse, forming
eventually the neutron star, is approximately
\begin{equation} 
  E_{bind} = (A m_{\rm 0} - M_{\rm g})  \quad,
\label{eq:binding}
\end{equation} 
where $m_{\rm 0}$ is defined as the mass per baryon of $^{56}$Fe. 
Analysis of data from
supernova 1987A leads to an estimate of $E_{\rm bind} = 3.8 \pm 1.2 \times
10^{53}$ erg~\cite{sch87}.

The gravitational red shift of the photons emitted radially
outwards from a neutron star surface is given by 
\begin{equation}
z_{surf} = \left( 1 - \frac{2GM_{\rm g}}{R}\right)^{-1/2} - 1
\label{eq:redshift}
\end{equation}
It is an important quantity needed, for example, for obtaining the mass 
and radius of a neutron star separately~\cite{lat05a}. 
Other quantities of interest for possible comparison with observational
data are the minimum rotation period, $\tau_{min}$~\cite{hes06}, and
the moment of inertia ${\mathcal I}$ (see Ref.~\cite{lat05b} and 
references therein). The minimum period is
given by the centrifugal balance condition for an equatorial fluid element
(i.e., the condition for it to be moving on a circular geodesic).
While determining this accurately requires using a numerical code for
constructing general-relativistic models of rapidly rotating stars, quite
good values can be obtained from results for non-rotating models using the
empirical formula~\cite{han89,han94}
\begin{equation}
\tau_{min}=0.820 \, \left( \frac{M^{\rm max}_{\rm g}}{M_\odot}\right)^{-1/2} 
\left( \frac{R_{\rm max}}{10\,{\rm km}}\right)^{3/2} \ {\rm ms} \, .
\label{eq:taumin}
\end{equation}
The shortest
period so far observed is 0.877 ms~\cite{hes06} but it is possible that
this limit may be connected with the techniques used for measuring pulsar
periods, rather than being a genuine physical limit. 

The moment of inertia $\mathcal I$
is calculated in this work by assuming that 
the rotation is sufficiently slow that the general relativistic {\it slow
rotation approximation}~\cite{har67} gives a rather accurate description
of the star and the surrounding space-time and using the expressions given 
in Ref.~\cite{rik02}. 

\section{Results and discussion}
\label{sec:resdis}
\subsection{Uniform matter}
\label{sec:unimat}

The properties of uniform matter and neutrons stars were calculated for 
12 EoS designed to systematically investigate the dependence on the basic 
QMC model parameters, summarised in Table~\ref{cap:Couplings}. 
There are three types of EoS used in this work. 
They have the same QMC parameters but differ by the limitation 
that we impose on the matter composition. The sets F-QMCx, 
where x refers to the  parametrisations of Table~\ref{cap:Couplings}, describe matter consisting of the full baryon octet plus electrons and muons. The sets N-QMCx  are used for $n+p+e+\mu$ matter and PNM-QMCx for pure neutron matter.      
For simplicity, we only show results for  QMC700 and QMC$\pi4$ in this section. 
These sets have the most realistic values of the QMC parameters and 
correspond to two rather different values of 
$K_\infty$, 340 and 256 {\rm\ MeV}. In the next section all the 
QMC parametrisations of Table~\ref{cap:Couplings} are used for the  
calculation of neutron star models and their properties are discussed 
in more detail.

In addition to the QMC EoS, we use four  other models for a comparison 
based on very different physics. The older Bethe-Johnson (BJ) 
EoS \cite{bet74} is based on the modified Reid potential and the 
medium effects are treated using constrained variational principle 
(model C in \cite{arn77}). The BJ EoS describes uniform matter including 
$n, p, \Lambda, \Sigma^{\rm \pm,0}, \Delta^{\rm \pm,0}$ 
and $ \Delta^{\rm ++}$.  The more recent EoS, calculated with the 
A18+$\delta v$+UIX$^*$ potential (APR)~\cite{akm98}, is based on
the Argonne A18 two-body interaction and includes three-body effects
through the Urbana UIX$^*$ potential. It also includes boost corrections to the
two-nucleon interaction which corresponds to the leading relativistic effects 
at order $v^2$. This interaction is considered one of the most
modern phenomenological potentials used for the description of   
nuclear matter containing nucleons only. As an example of a phenomenological,   
nucleon-only EoS the Skyrme SkM$^*$ interaction is used, in particular 
because of the similarity of its results for finite nuclei to those of 
the effective force derived from the QMC model~\cite{gui04,gui06}. Finally, 
we also compare with the EoS of a hybrid (neutron-quark) 
star matter of Glendenning 
(Table 9.1 of \cite{gle00}) (Hybrid)  calculated in RMF approximation.  
%
%
\begin{figure}
\centerline{\includegraphics[clip,angle=0,width=11cm,height=7cm]{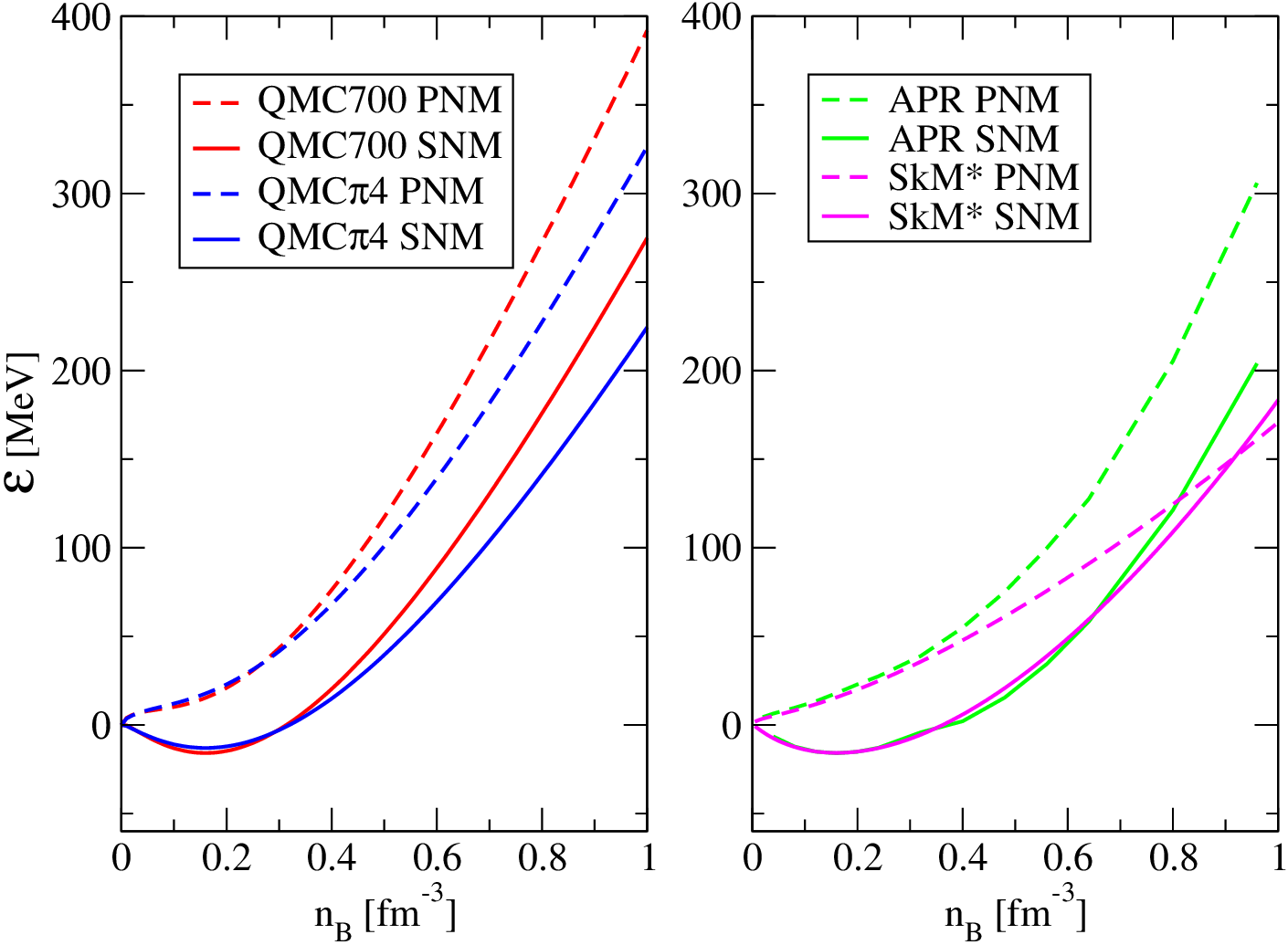}}
\caption{\label{fig:pnsym} The energy per particle for pure neutron and 
symmetric nuclear matter is plotted as
a function of baryon number density $n_B$ for the QMC700 and QMC$\pi4$ 
interactions (left panel). For comparison, also shown are equivalent curves 
calculated using the APR potential~\protect\cite{akm98} and the 
Skyrme SkM$^*$ parametrisations (right panel)}
\end{figure}
The  binding energy per particle in pure neutron matter (PNM) and 
symmetric nuclear matter (SNM), for QMC700 and QMC$\pi4$, are illustrated 
in Fig.~\ref{fig:pnsym} in comparison with the APR and SkM$^*$ predictions. 
At lower densities the calculated values are very similar. However, 
with increasing density the differences become more dramatic. In particular, 
while the QMC and APR models predict a steady increase of the energy per 
particle in both pure neutron and symmetric matter at about the same rate, 
the SkM$^*$ model yields a more rapid growth of the energy per particle in 
SNM than in PNM, leading to crossing at about 0.9{\rm\ fm}$^{\rm -3}$. 
As shown in Fig.~\ref{fig:as} and discussed in detail in \cite{rik03}, 
this leads to negative symmetry energy and a collapse of nuclear matter. 
Although the QMC and Skyrme energy functionals have similar structure and 
yield similar results for finite nuclei at normal nuclear density, 
QMC has a different density and isospin dependence~\cite{gui06} 
that improves the EoS at high densities in comparison with SkM$^*$, 
while the symmetry energy shows a slower increase with density than 
the APR model. There is no experimental evidence concerning the density 
dependence of the symmetry energy, as it is not possible to study 
the properties of uniform matter at high densities in terrestrial conditions.
Even in heavy-ion collisions at high energies the maximum 
baryon density reached  
is unlikely to exceed $2\div 3 n_0$ in the phase of the reaction closest to the conditions in cold neutron stars. 
Nevertheless, there is some indirect evidence which seems to support 
models which predict that the symmetry energy grows with 
density \cite{rik03}. Such models yield sensible results for properties 
of neutron stars that can be compared with observations 
(see Section~\ref{sec:nstars}).    
\begin{figure}
\centerline{\includegraphics[clip,angle=0,width=9cm,height=6cm]{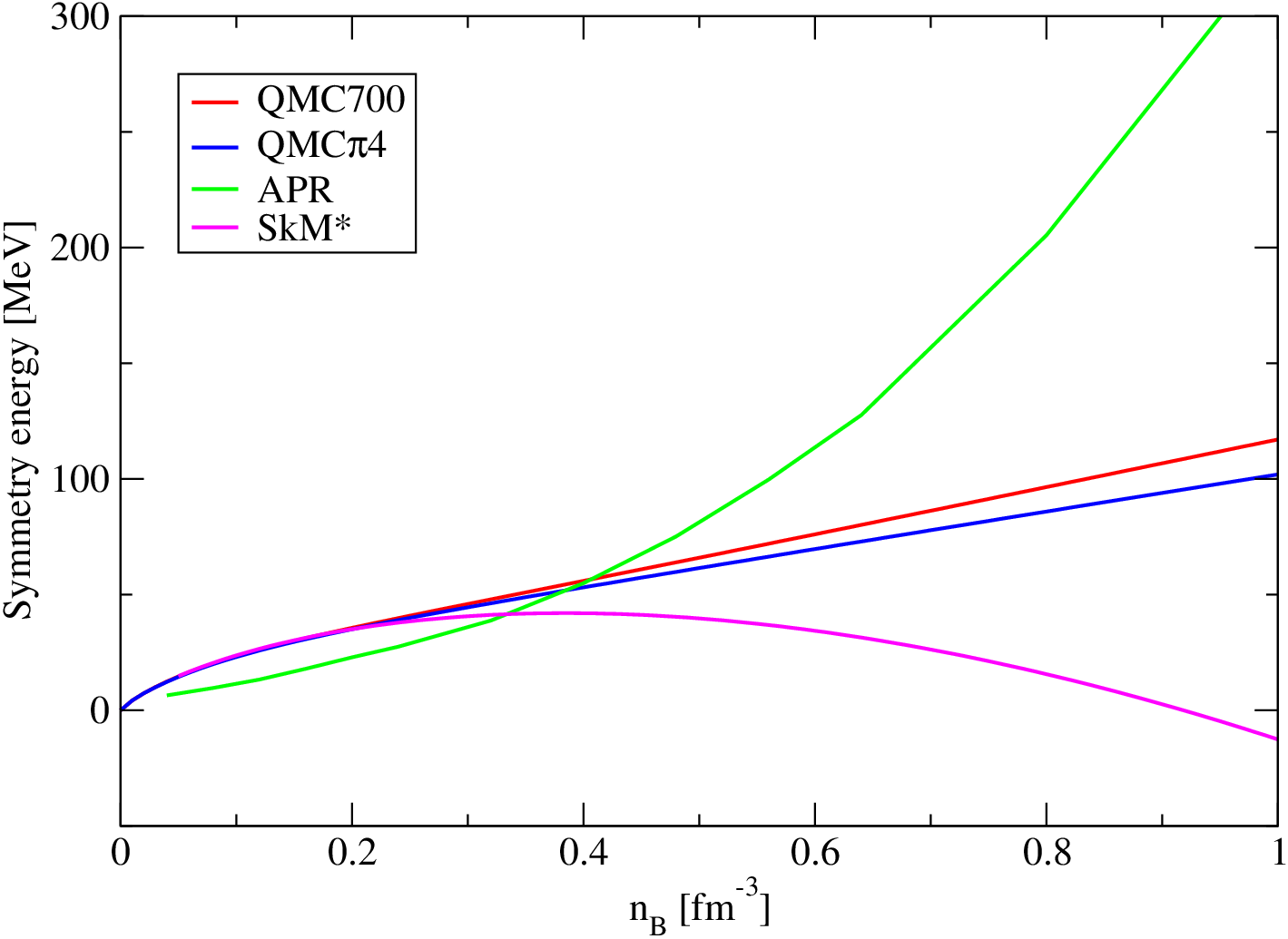}}
\caption{\label{fig:as} Symmetry energy as
a function of baryon number density $n_B$ for QMC700 and QMC$\pi4$. 
Also shown are the corresponding results for the APR potential and the 
Skyrme SkM$^*$ parametrisations.}
\end{figure}

\begin{figure}
\centerline{\includegraphics[clip,angle=0,width=11cm,height=7cm]{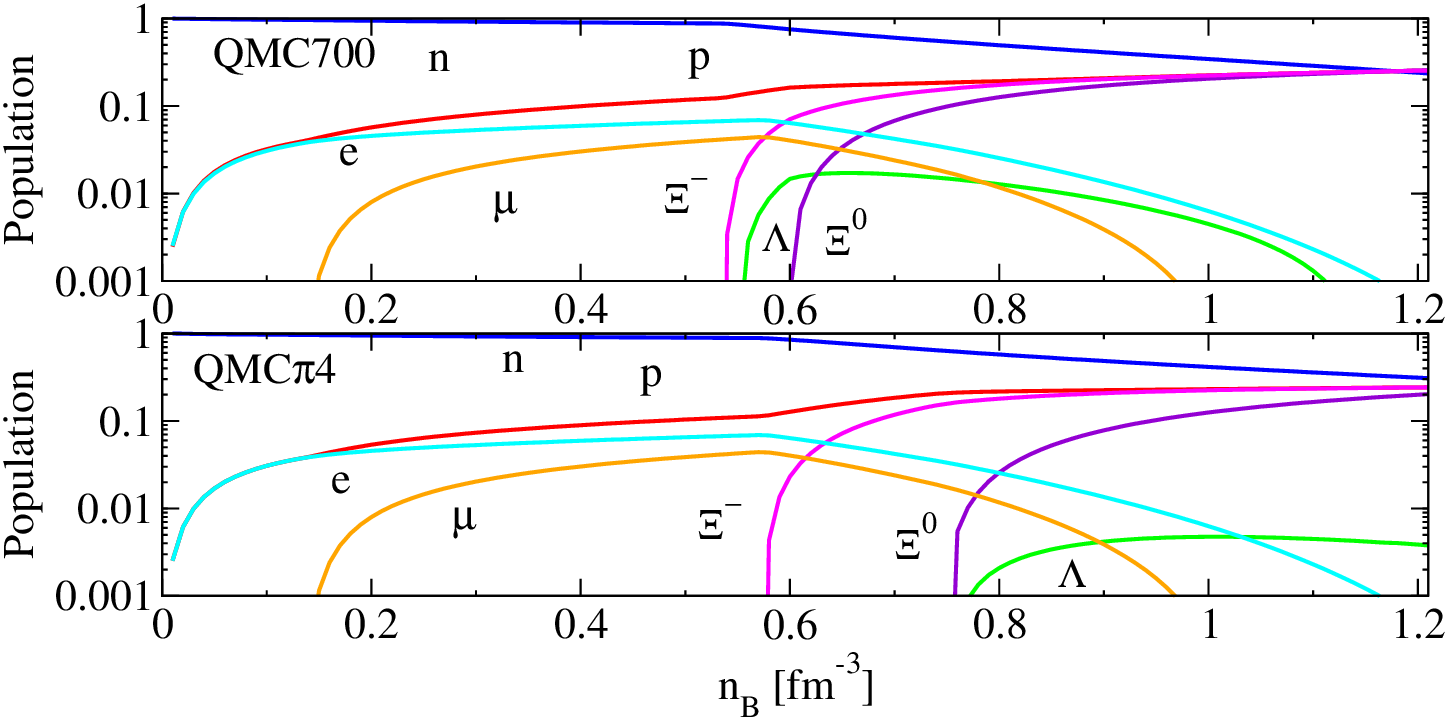}}
\caption{\label{fig:equilibrium} Relative population of baryons and leptons 
in $\beta$-equilibrium matter for QMC700 and QMC$\pi4$. For more 
explanation see text.}
\end{figure}

The results of our calculation of the equilibrium composition of uniform 
matter for QMC700 and QMC$\pi4$  are shown in Fig.~\ref{fig:equilibrium}. 
In a striking difference from all other models which include the full 
baryon octet, QMC does not predict the production of hyperons at 
densities less than 3 times nuclear saturation density. Furthermore, 
the first hyperons to appear are the cascades, $\Xi$, together with the 
$\Lambda$ hyperon. The $\Sigma$ hyperons are not produced at densities 
below 1.2\ {\rm\ fm}$^{-3}$. 
This scenario is a direct consequence of three factors which are present 
in the QMC model and absent in the others. First, we use the Hartree-Fock 
approximation, while the earlier calculations worked only at the Hartree 
level. Actually, we see no reason to ignore antisymmetrisation, which is  
the key to our successful interpretation ofthe spectra of finite 
nuclei~\cite{gui06}. Second, the color hyperfine interaction, which is 
necessary to explain the mass splitting within the octet, as well as the 
$N \Delta$ splitting, induces a flavor dependent effective mass which 
strongly disfavors the $\Sigma$. Third, the scalar polarisability, $d$, 
induces many body effects which tend to amplify the previous ones. 

To make these considerations 
quantitative, we have computed the single 
particle threshold densities on top of neutron matter, using the QMC700 
parametrization and the variations on it defined earlier.  The results 
are shown in Table~\ref{cap:Thresholds}, where the first column corresponds 
to the full  QMC700 model as shown in Fig.~\ref{fig:equilibrium}. In the 
second column we show the same calculation but with the hyperfine color 
interaction and the scalar polarisability set to zero for the strange 
particles. Clearly this produces a very spectacular effect on the 
$\Sigma$ thresholds, which become comparable to the $\Xi$ thresholds. 
In the third column we show the same calculation as in the first but now in  
the Hartree approximation, that is dropping the exchange terms in 
Eqs.~(\ref{eq:QMC14}-\ref{eq:QMC17}) and readjusting the coupling 
to reproduce the saturation point of nuclear matter. One observes again 
a strong rearrangement of the threshholds in favor of the $\Sigma$. Finally, 
the last column shows the combination of both effects. With these 
three alterations, the QMC model obviously  becomes similar to the SU(3)  
relativistic mean field models -- with respect to strangeness production. 
In particular, the first threshold is now for the $\Sigma^-$, almost 
identical to that for the $\Lambda$ and it occurs at less than twice 
the nuclear density.

\begin{figure}
\centerline{\includegraphics[clip,angle=0,width=11cm,height=7cm]{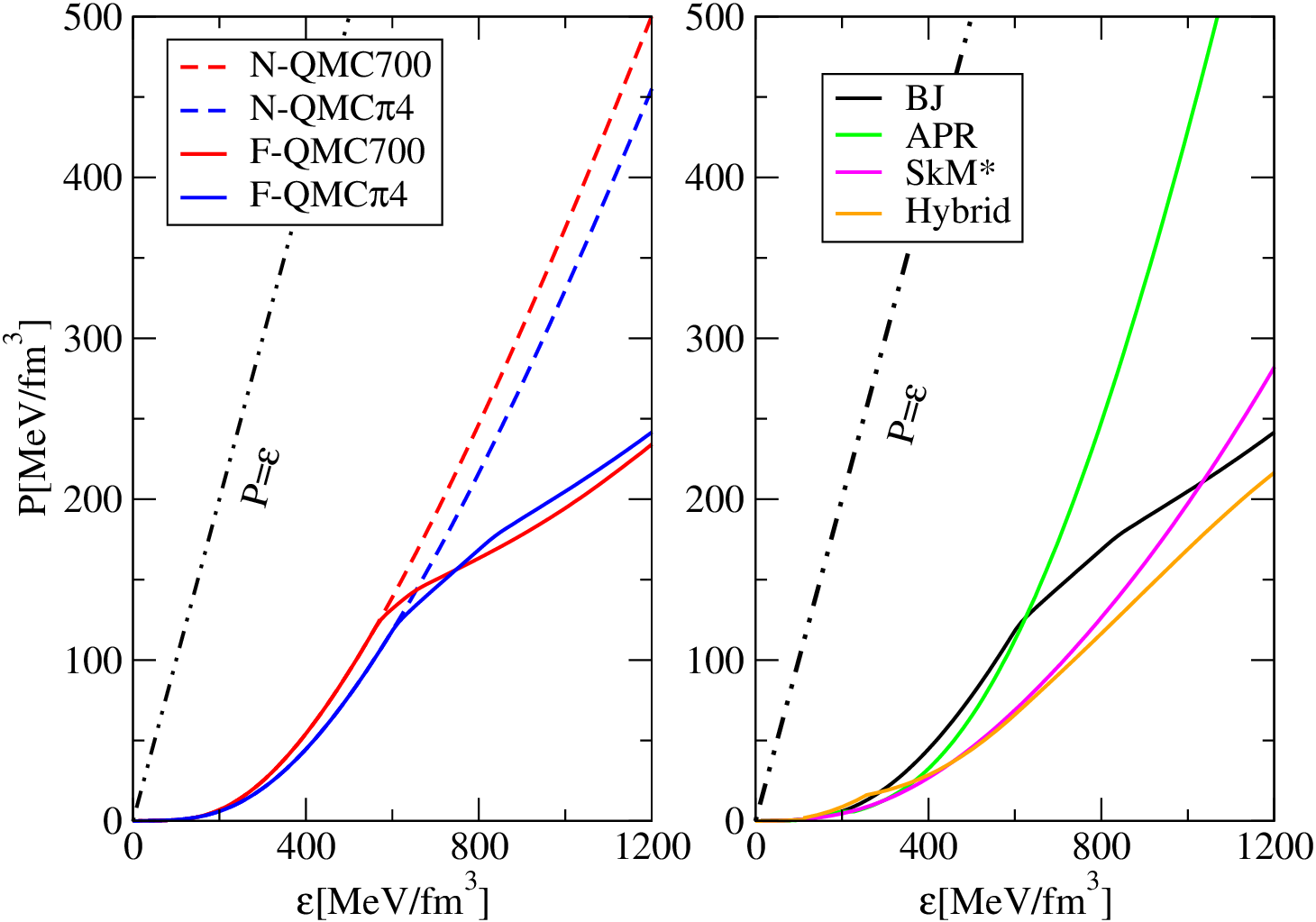}}
\caption{\label{fig:pe} Left panel: Pressure as a function of total energy density for sample EoS for $\beta$-equilibrium matter containing  nucleons p+n+e+$\mu$  (N-QMC700, N-QMC$\pi$4)  and the full octet  (F-QMC700, F-QMC$\pi$4). Right panel: Results of BJ, APR, SkM$^*$ and hybrid EoS. The curve $P=\epsilon$  shows the stiffest possible EoS.}
\end{figure}

\begin{table}
\begin{center}\begin{tabular}{|c|c|c|c|c|}
\hline 
&
Hartree-Fock&
Hartree-Fock&
Hartree&
Hartree\tabularnewline
&
QMC700&
QMC700, hfc=d=0&
QMC700&
QMC700, hfc=d=0\tabularnewline
\hline
\hline 
$\Lambda$&
0.555&
0.42&
0.35&
0.31\tabularnewline
\hline 
$\Sigma^{-}$&
0.92&
0.51&
0.41&
0.3\tabularnewline
\hline 
$\Sigma^{0}$&
0.97&
0.58&
0.61&
0.44\tabularnewline
\hline 
$\Sigma^{+}$&
1.01&
0.63&
0.87&
0.61\tabularnewline
\hline 
$\Xi^{0}$&
0.59&
0.54&
0.58&
0.55\tabularnewline
\hline 
$\Xi^{-}$&
0.55&
0.50&
0.45&
0.43\tabularnewline
\hline
\end{tabular}\end{center}
\caption{\label{cap:Thresholds}Single particle threshold densities 
(in fm$^{-3}$)  on top of neutron
matter for QMC700, either in Hartree or Hartree-Fock approximation
and with the hyperfine color interaction (hfc) and scalar polarisability, $d$,
set to zero in the strange sector.}
\end{table}
%
%
As expected, the presence of hyperons does soften the EoS, as demonstrated 
in Fig.~\ref{fig:pe}. We observe that  QMC700 and QMC$\pi4$ both  
show a very similar EoS to that for the BJ model, which also contains 
hyperons. However, it is clearly demonstrated that the composition of 
the stellar matter is not the only way to soften the EoS. The pressure 
also depends critically on the potentials acting between the particles 
present and, in turn, on the density dependence of the symmetry energy. 
Thus, for example, the SkM$^*$ Skyrme model produces a very soft EoS, 
in contrast to APR, and almost the same EoS as the neutron+quark RMF hybrid 
model. Clearly more constraints are needed to distinguish amongst these 
possibilities. Further illustrations of the variety of possibilities for the 
dependence of the pressure on the total energy density can be found 
in Ref.~\cite{pag06}.
 
To conclude this Section, let us mention that we have also studied the 
possibility that matter at high density may contain exotic particles, 
such as pentaquarks or six quarks bags. The calculation goes along the 
same lines as for the ordinary baryonic matter and poses no conceptual 
difficulties because, in beta equilibrium, one does not worry about the exact 
way a particular component has been produced. The practical problem is 
that these would-be exotics have not been identified as free particles 
and therefore their free masses (if they exist) are  essentially unknown. 
To get an estimate of these masses, which are of course the critical 
parameters,  we must rely on the bag model. It turns out that, even if we 
allow the exotics to be lighter  than the bag model prediction by 
as much as 400 MeV, they do not appear at any significant level when 
the conditions for beta equilibrium are imposed on matter where the full 
octet is allowed.

\begin{table}
\begin{center}
\begin{tabular}{|c|c|c|c|c|c|c|}
\hline 
&
$N_{1}$&
$p_{1}$&
$N_{2}$&
$p_{2}$&
$r$&
$a$\tabularnewline
\hline
\hline 
N-QMC700&
0&
&
8.623~10$^{-3}$&
1.548&
342.4&
184.4\tabularnewline
\hline 
N-QMC$\pi4$&
0&
&
4.634~10$^{-3}$&
1.623&
346.7&
192.1\tabularnewline
\hline 
N-QMC700&
2.62~10~$^{-7}$&
3.197&
0.0251&
1.286&
522.1&
113\tabularnewline
\hline 
F-QMC$\pi4$&
8.476~10$^{-8}$&
3.328&
0.03178&
1.254&
514.5&
145.5\tabularnewline
\hline
\end{tabular}
\end{center}

\caption{\label{cap:EOSFit}The parameters of the fit defined by eq.\ref{eq:eosfit}}
\end{table}

For the readers convenience we have preformed an analytical fit of
our EOS for the cases shown on Figure \ref{fig:pe}. It has the form:
\begin{equation}
P=\frac{N_{1}\varepsilon^{p_{1}}}{1+\exp[(\varepsilon-r)/a]}+\frac{N_{2}\varepsilon^{p_{2}}}{1+\exp[-(\varepsilon-r)/a]}\label{eq:eosfit}.
\end{equation}
The parameters $(N_{1},p_{1},N_{2},p_{2},r,a)$ are given in Table
\ref{cap:EOSFit} with $(P,\varepsilon)$ expressed in $MeV/fm^{3}$. The fit is accurate in the range $\rho_B=0\div1.2fm^{-3}$.

\subsection{Neutron Stars}
\label{sec:ns}

The calculated properties of neutron stars with the maximum mass for 
all 12 model EoS are summarised in Table~\ref{tab:starmax}. For comparison, 
the results for the BJ, APR, SkM$^*$ and Hybrid EoS are also included. 
The same information is given for these models for stars 
with `canonical' mass 1.4 M$_\odot$ 
in Table~\ref{tab:star14}. It should be noted that the entries 
in Table~\ref{tab:star14} do not differ for the full octet and nucleon-only 
versions of the model, 
because the central density of 1.4 M$_\odot$ stars is below 
the threshold for the creation of hyperons.   

\begin{table}
\caption{\label{tab:starmax} Parameters of maximum mass, cold neutron stars 
for the full set of QMC EoS. $n_c$ and $\rho_c$ are the central 
baryon number density [{\rm\ fm}$^{-3}$] and total mass
density [$10^{15}$ g cm$^{-3}$] respectively, $R$ [km] is the radius, 
the maximum gravitational mass is M$_\odot$, $A$
[10$^{57}$] is the total baryon number, $E_{\rm bind}$ [10$^{53}$ erg] is
the binding energy and $\tau$ [msec] is the estimate of the rotation
period calculated from Eq.~(\ref{eq:taumin}), $v_{\rm s}$ is the speed of 
sound and $z_{\rm surf}$ is the radiational red-shift. 
The index `max' has been dropped for all observables in the 
header of this table.}
\vskip 10pt 
\begin{tabular}{r|ccccccccc}
EoS  &  $n_{\rm c}$  & $\rho_{\rm c}$& $R$ & $M_{\rm g}$ & $A$ & $E_{\rm bin}$ &$\tau$ & $v_{\rm s}$ & $z_{\rm surf}$\\ \hline
  F-QMC600      & 0.81 &  1.66 & 12.45 &  1.99 &  2.75 &  5.85  &  0.81  & 0.65  & 0.38 \\
  F-QMC700      & 0.82 &  1.68 & 12.38 &  1.98 &  2.74 &  5.85  &  0.80  & 0.65  & 0.38 \\
  F-QMC$\pi1$      & 0.85 &  1.75 & 12.19 &  1.96 &  2.72 &  5.80  &  0.79  & 0.66  & 0.38 \\
  F-QMC$\pi2$      & 0.86 &  1.79 & 12.08 &  1.93 &  2.66 &  5.56  &  0.78  & 0.65  & 0.38 \\
  F-QMC$\pi3$      & 0.89 &  1.85 & 11.93 &  1.90 &  2.62 &  5.42  &  0.77  & 0.66  & 0.37 \\
  F-QMC$\pi4$      & 0.93 &  1.93 & 11.70 &  1.85 &  2.54 &  5.17  &  0.76  & 0.66  & 0.37 \\
  N-QMC600  & 0.96 &  2.18 & 11.38 &  2.22 &  3.14 &  7.68 &  0.67  & 0.84  &0.54 \\
  N-QMC700  & 0.96 &  2.19 & 11.34 &  2.21 &  3.14 &  7.69 &  0.67  & 0.84  &0.54 \\
  N-QMC$\pi1$  & 0.99 &  2.25 & 11.19 &  2.18 &  3.09 &  7.55 &  0.66  & 0.84  &0.55 \\
  N-QMC$\pi2$  & 1.01 &  2.31 & 11.08 &  2.15 &  3.04 &  7.31 &  0.65  & 0.84  &0.55 \\
  N-QMC$\pi3$  & 1.04 &  2.38 & 10.94 &  2.12 &  2.99 &  7.13 &  0.64  & 0.84  &0.53 \\
  N-QMC$\pi4$  & 1.09 &  2.49 & 10.73 &  2.07 &  2.91 &  6.84 &  0.63  & 0.84  &0.53 \\
  \hline
  BJ         & 1.31 &  3.04 &  9.92 &  1.851 &  2.56 &  5.41 &  0.60  &&0.49 \\
  APR        & 1.15 &  2.76 &  9.99 &  2.201 &  3.21 &  9.03 &  0.55  &$>$1.00 & 0.69 \\
  SkM$^*$    & 1.66 &  3.83 &  8.94 &  1.617 &  2.22 &  4.47 &  0.55  & 0.82  &0.46 \\
  Hybrid     & 1.33 &  2.81 & 10.42 &  1.453 &  1.93 &  3.02 &  0.72  &&0.30 \\ \hline
\end{tabular}
\end{table}
\begin{table}
\caption{\label{tab:star14} The same as Table~\protect\ref{tab:starmax} 
but for 1.4 M$_\odot$ star models. Data for N-QMCx EoS are not shown as 
they are identical to results of F-QMCx EoS because the central density 
of all 1.4 M$_\odot$ stars is predicted to be below the threshold for the   
appearance of hyperons.}
\begin{tabular}{r|ccccccccc}
EoS  &  $n_{\rm c}$  & $\rho_{\rm c}$& $R$ & $A$ & $E_{\rm bin}$ &$\tau$& $v_{\rm s}$ & $z_{\rm surf}$\\ \hline
  F-QMC600     &  0.39 &  0.69 & 12.94 &  1.86 &  2.76 &  1.02 & 0.58 & 0.21 \\
  F-QMC700     &  0.39 &  0.70 & 12.88 &  1.86 &  2.76 &  1.01  & 0.58 & 0.21\\
  F-QMC$\pi1$     &  0.41 &  0.73 & 12.74 &  1.86 &  2.79 &  0.99 & 0.58 & 0.22\\
  F-QMC$\pi2$     &  0.42 &  0.75 & 12.67 &  1.85 &  2.75 &  0.99 & 0.59& 0.22\\
  F-QMC$\pi3$     &  0.43 &  0.78 & 12.55 &  1.85 &  2.75 &  0.97 & 0.59& 0.22\\
  F-QMC$\pi4$     &  0.46 &  0.83 & 12.38 &  1.86 &  2.79 &  0.95 & 0.59& 0.28\\
  \hline
  BJ        &  0.59 &  1.09 & 11.86 &  1.86 &  2.76 &  0.89 & & 0.24\\ 
  APR       &  0.55 &  1.00 & 11.47 &  1.87 &  3.03 &  0.85 & & 0.29\\
  SkM$^*$       &  0.87 &  1.68 & 10.51 &  1.88 &  3.12 &  0.74 & 0.62 & 0.25\\
  Hybrid    &  0.91 &  1.79 & 11.20 &  1.86 &  2.76 &  0.82 & & 0.28\\ \hline
\end{tabular}
\end{table}
\begin{figure}
\centerline{\includegraphics[clip,angle=0,width=11cm,height=7cm]{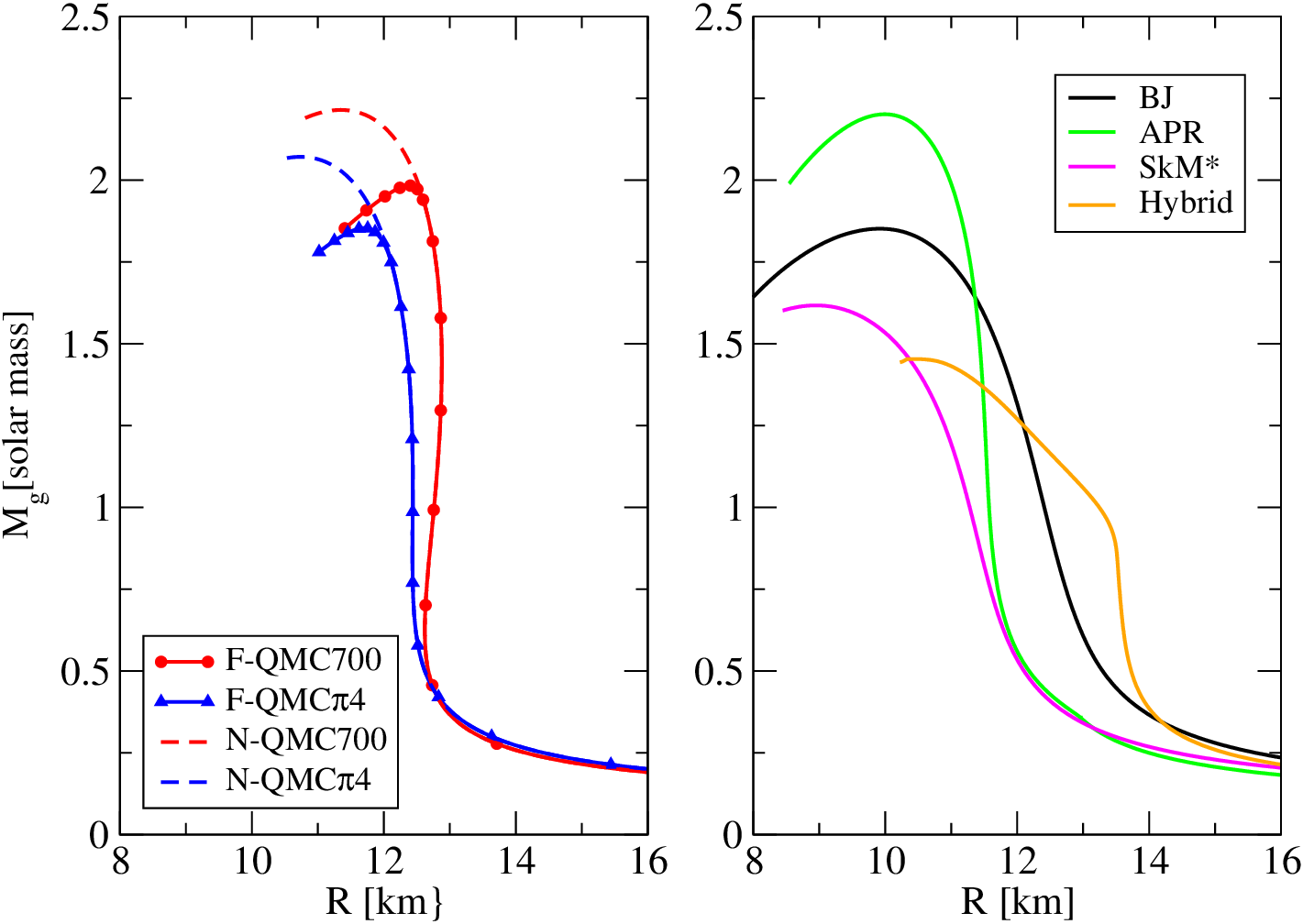}}
\caption{\label{fig:massradius}The gravitational masses of non-rotating
neutron-star models (measured in solar masses) plotted against radius (in
kilometers), calculated for selected QMC EoS -- see the text for more 
explanation.
}
\end{figure}
The relation between the gravitational mass and radius for various neutron 
star models is shown in Fig.~\ref{fig:massradius}, for a representative 
set of EoS, chosen to be F-QMC700, F-QMC$\pi4$, N-QMC700 and N-QMC$\pi4$.    

Apart from the basic mass-radius relation, there are other features of
neutron stars that provide the possibility to constrain the EoS
and thus the nucleon-nucleon interaction in stellar matter. The volume and 
quality  of the observational data has increased in recent years, 
offering more options for testing the physical basis of the EoS.
\begin{figure}[h]
\centerline{\includegraphics[clip,width=9cm,height=6cm]{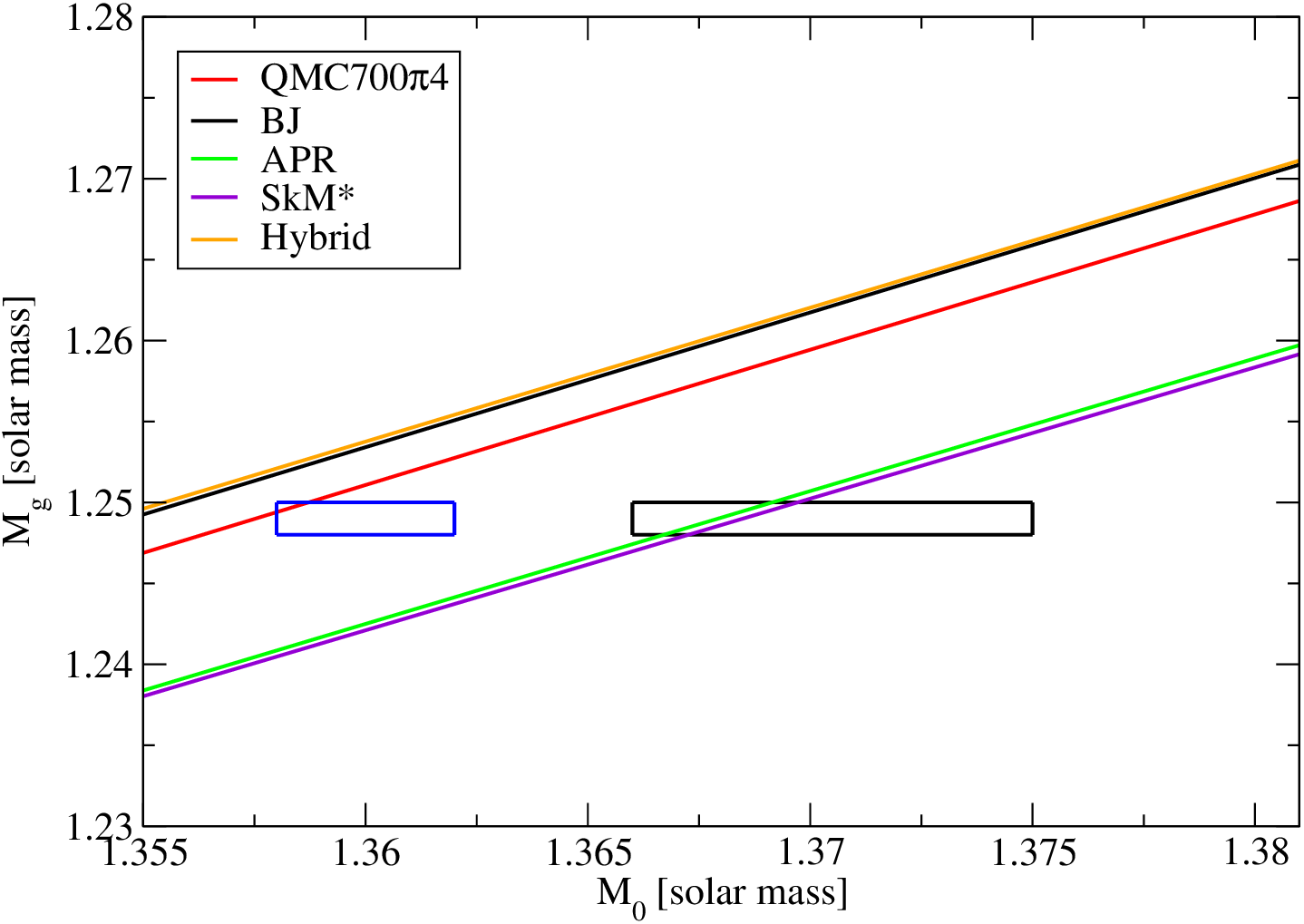}}
\caption{\label{fig:lowmass}Relation between the gravitational mass, 
M$_{\rm g}$, for various 
neutron star models (including the QMC EoS) and the corresponding 
baryonic mass, M$_{\rm 0}$. The boxes represent constraints derived by Podsiadlowski et al.\cite{pod05} (black box) and more recently by   Kitaura et al.\cite{kit06} (blue box), based on the properties os system J0737-3039, as discussed in the text.}   
\end{figure}
An important constraint on the EoS at relatively low densities has 
been recently identified~\cite{pod05}
in connection with the very precise determination of the gravitational
mass of Pulsar B in the system J0737-3039, $M_{\rm
g}$~=~1.249$\pm$0.001 $M_\odot$. If the progenitor star was a white dwarf with O-Ne-M core and Pulsar B was  formed in
an electron-capture supernova, a possibility supported in part by new observations \cite{sta06}, a rather narrow range for the baryonic
mass, $M_{\rm 0}$, between 1.366--1.375 $M_\odot$,   
can be determined for the pre-collapse core. Assuming that 
there is no (or negligible) baryon loss during the
collapse, the newly born neutron star will have the same baryonic mass
as the progenitor star. For any given EoS for cold neutron star matter the
relation between the gravitational and baryonic mass can be 
calculated and tested against a very narrow window (full black line box in Fig.~\ref{fig:lowmass})defined by the data for Pulsar B. The width of the window reflects uncertainty in modelling the composition of the progenitor core, namely electron-capture rates, nuclear network calculation, Coulomb and general relativity corections. Very recently, another simulation of the  explosion of O-Ne-Mg cores \cite{kit06} has been performed  which predicts the baryonic mass of Pulsar B 1.36$\pm$0.002 $M_\odot$ (full blue line box in Fig.~\ref{fig:lowmass}). The error includes the uncertainty in the EoS used and the wind ablation after the simulation was terminated. This model (see Ref.~\cite{kit06}) includes a small mass loss between 0.014 to 0.017 $M_\odot$ in the process. As seen in Fig.~\ref{fig:lowmass}, F-QMC$\pi$4 (the other versions give essentially the same curves) agrees with the latter constraint and predicts mass loss in the region of 0.008 - 0.017 $M_\odot$ in the model of Podsiadlowski et al \cite{pod05}, which is rather similar to the result of Kitaura et al \cite{kit06}. The other sample EoSs either do not predict any mass loss (APR and SkM$^*$) or they predict  a mass loss which is too large.        

%
\begin{figure}
\centerline{\includegraphics[clip,angle=0,width=9cm,height=6cm]{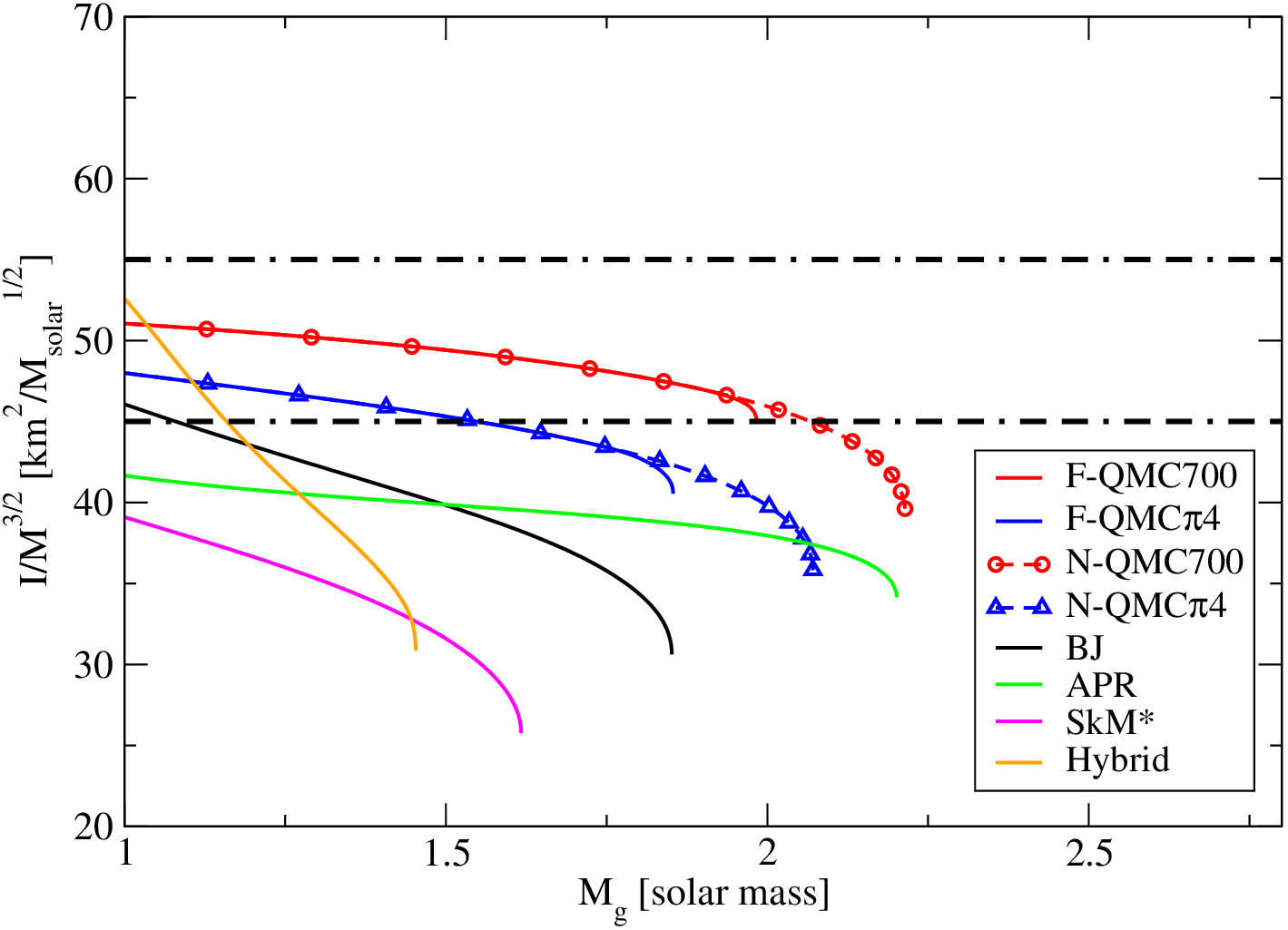}}
\caption{\label{fig:lat1} Moment of inertia, scaled by $M^{\rm 3/2}$,  
as a function of the stellar mass, $M$, for sample QMC EoS. 
The BJ, APR, SkM$^*$ and Hybrid EoS results are added for comparison. 
The dash-dot lines indicate the predicted experimental constraints from 
Ref.~\cite{lat05b} - see the text for details.}
\end{figure}
%

Another interesting possibility has been recently suggested by Lattimer and
Schultz \cite{lat05b}. Measurement of the moment of inertia of 
Pulsar A (with known mass of about 1.34 M$_\odot$) in the same system, to $\sim$10\% precision, would allow a rather accurate estimate of
the radius of the star and of the pressure in matter of density in the
vicinity of $1\div2 n_{\rm 0}$. Such information could provide a rather strong
constraint on the EoS of neutron stars. In Fig.~\ref{fig:lat1}, using exactly the same scale as as Fig.~1 of Ref.~\cite{lat05b}, we illustrate this possibility. The dashed horizontal lines depict the error range of a hypothetical result of a measurement, selected as close as possible to the expected value. If this were the actual experimental result, many EoS would not be useful in determining  the radius of Pulsar A, as they would cross the area between the two dashed lines either below or above M = 1.34 M$_\odot$. The QMC EoS would satisfy this constraint. Moreover, their relatively weak mass dependence would be an advantage in the study of of another analogous system with known mass up to $\sim$1.6 M$_\odot$ (N-QMC$\pi$4) and $\sim$2 M$_\odot$. We note that incorporation of hyperons shortens this range at higher masses, as illustrated in Fig.~\ref{fig:lat1}. 

\begin{figure}[t]
\centerline{\includegraphics[clip,angle=0,width=9cm,height=6cm]{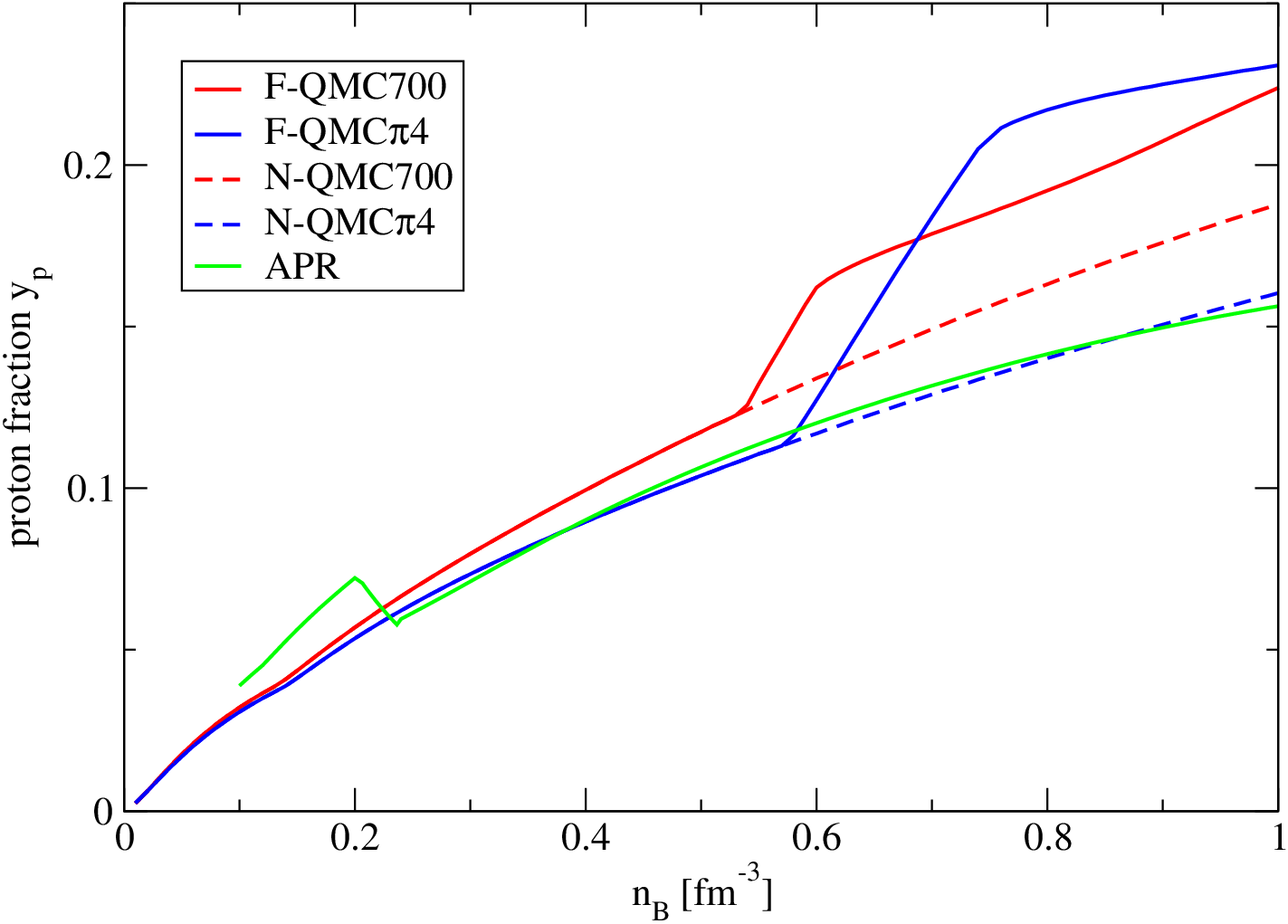}}
\caption{\label{fig:yp} The proton fraction $y_{\rm p}$ for beta stable 
matter, as calculated using the QMC EoS
is plotted as a function of the baryon number density $n_B$. 
The data for the APR EoS are added for comparison. }
\end{figure}

Finally, we study the implications of the baryon composition of stellar 
matter, as calculated using the sample EoS, for a possible cooling mechanism 
of neutron stars either just after their
birth in supernovae or after heating in an accretion episode. 
The cooling mechanisms can, in principle,
provide further important constraints on neutron-star models. 
However,
as discussed in more detail previously~\cite{rik02}, the cooling
processes for both young and old neutron stars are not currently known
with any certainty, although several scenarios have been proposed
\cite{pag04,jon01}. Many of them involve emission of neutrinos from 
the stellar core. The most frequently discussed, the direct URCA process, 
produces neutrinos in proton-neutron weak decays with an additional production of either electrons of muons. The proton concentration in the matter 
is a crucial ingredient of the process. Conservation of energy and momentum 
requires %
$y^{1/3}_{\rm n} < y^{1/3}_{\rm p} + y^{1/3}_{\rm e}$, or  
$y^{1/3}_{\rm n} < y^{1/3}_{\rm p} + y^{1/3}_{\rm \mu}$, to be satisfied. 
This, in turn, can happen only in matter where the proton fraction grows 
steadily with increasing density and $y_p$ is greater than about 0.11 
for matter consisting of only protons, neutrons and electrons. 
Medium effects and additional interactions among the particles change 
this number only slightly~\cite{pag06} but the presence of muons increases 
the critical ratio, $y_p$, to about 0.15. 
We display the proton fraction for 
sample EoS in Fig.~\ref{fig:yp}.
It is clear that the proton fraction in F-QMC700 and F-QMC$\pi 4$ pass this critical 
limit at densities close to $0.6 {\rm\ fm}^{-3}$, coincident with the 
appearence of negative $\Xi^-$ hyperons. This density is well below the 
central density for maximum mass models and thus the direct URCA process 
is allowed for models of astrophysical interest, in agreement with expectation. On the contrary,  
for the N-QMC700 and N-QMC$\pi 4$ models the critical density for the  start of the direct URCA process is much higher, 
between $0.8\div 1.0{\rm\ fm}^{-3}$, and is close to or exceeds the 
inferred maximum central density, suggesting that this cooling mechanism 
would not be allowed in these cases. This reinforces our message concerning the importance of incorporating of the full baryon octet in QMC.

\section{Conclusions}
\label{sec:concl}
One of the most important features of the QMC EoS is that although the 
QMC and Skyrme energy functionals have similar structure at densities 
close to nuclear saturation density, they differ significantly at higher 
densities, because of their different density and isospin dependence. 
The QMC EoS provide cold neutron star models with maximum mass 
1.9--2.1 M$_\odot$, with central density less than 6 times nuclear 
saturation density and thus offer a consistent description of the stellar 
mass up to this density limit, without extrapolation beyond the region of 
validity -- as needed, for example, when the non-relativistic, 
nucleon-only Skyrme models 
are used at these densities. The maximum mass of neutron stars in the 
QMC model is close to the most recent observation, the central density 
is relatively low in comparison to other models and all the predicted 
stars are well within the causal limit. 

The QMC model includes self-consistently the presence of hyperons 
at high densities.
Unlike most other models, no hyperon contribution at densities lower 
than $3\,n_0$ is predicted in $\beta$-equilibrium matter. At higher densities, 
$\Xi^{-,0}$ and $\Lambda$ hyperons are present. The absence of the 
$\Sigma^{\rm \pm,0}$ hyperons is due, on the one hand, to the unique features 
of the model, namely the implementation of the color hyperfine interaction 
and the scalar polarisability of the baryons and, on the other hand, to the 
use of the Hartree-Fock approximation. The pronounced effects of the 
presence of hyperons on the properties of dense matter and neutron star models 
have been identified by a comparison of nucleon-only EoS 
with full-baryon-octet EoS.    
In nuclear matter, the presence of hyperons, as expected, softens the EoS 
as illustrated in Fig.~\ref{fig:pe}. However, it needs to be stressed 
that the composition of stellar matter at high density is not necessarily 
the only determining factor of the behaviour of pressure with increasing 
density. In fact, it is critically dependent on 
the potentials acting amongst particles 
present in the matter. 

In neutron star models, the hyperon component 
systematically reduces the prediction of the maximum mass of a neutron star and 
increases its radius at maximum mass, as demonstrated 
in Fig.~\ref{fig:massradius}. This is a very important finding, because 
many non-relativistic EoS, including those based on the effective Skyrme 
interaction and the A18+$\delta$v+UIX$^*$ potential, do not include 
hyperons at 
densities corresponding to the central density of maximum mass star 
models and thus their results should be taken with caution. Moreover, 
the relationship between the moment of inertia and the stellar gravitational 
mass, $M_{\rm g}$, is clearly influenced (see Fig.~\ref{fig:lat1}), as 
is the cooling mechanism expected to take place in neutron stars - 
the direct URCA process - is predicted to be allowed in stars with hyperons 
but not in nucleon-only neutron stars.
It is also important to realise that the models for the nucleon-hyperon and 
hyperon-hyperon interactions play an important role in this comparison.
In future, it would be interesting to apply the present QMC model to 
non-equilibrium matter at finite temperature and produce EoS that could 
be tested in core-collapse supernova simulation models. The high density 
threshold for the appearance of hyperons probably rules out any 
hyperon-related effects taking place in the collapsing star at the 
bounce density of $1.5 - 2.0\,n_0$, in variance with some other EoS 
with a much lower threshold density (e.g. \cite{men04}).   

\section{Acknowledgement}
This work has been supported by US DOE grant
DE-FG02-94ER40834 and by US DEO grant DE-AC05-06OR23177, under 
which Jefferson Science Associates operates Jefferson Lab and by the {\it Espace de Structure Nucl\'eaire Th\'eorique du CEA}. (JRS) wishes to acknowledge helpful discussions with P.~Podsiadlowski and J.~C.~Miller and to  thank J.~C.~Miller for providing a code for calculation of properties of slowly rotating neutron stars.

\section{Appendix}

\subsection{Free mass of the baryons in the bag model}

The free mass of the baryon of flavor $f$ with quark content $(N_{ud},N_{s})$
is written as\begin{equation}
M_{f}=\frac{N_{ud}\Omega(m_{ud})+N_{s}\Omega(m_{s})}{R}-\frac{Z_{0}}{R}+\Delta E_{M}(f)+\frac{4\pi BR^3}{3}\label{eq:A1}\end{equation}
 where the quark mode $\Omega(m)$ is determined by the boundary condition

\begin{equation}
\frac{\sin x}{x}+\frac{1}{\Omega+mR}\left(\cos x-\frac{\sin x}{x}\right)=0,\,\,\,\, x=\sqrt{\Omega^{2}-(mR)^{2}}\label{eq:A2}\end{equation}
and $B$ is the bag constant. In the presence of the scalar field
one can have \textbf{$\Omega^{2}-(mR)^{2}<0$} but the equation remains
valid by analytical continuation. The bag radius is fixed for each
flavour by the stability condition\begin{equation}
\frac{\partial M_{f}}{\partial R}=0.\label{eq:A3}\end{equation}
 We set the $u,d$ mass equal to zero and the zero point parameter
$Z_{0}$ is assumed to be the same for all particles. For completeness
we briefly recall how the hyperfine color 
interaction $\Delta E_{M}$ is evaluated
according to~\cite{DeGrand:1975cf}\footnote{We thank H. Grigorian who pointed out a missprint in Equation 2.23 of~Ref.\cite{DeGrand:1975cf}}. For a color singlet baryon one has
\begin{equation}
\Delta E_{M}=\sum_{i<j}E_{ij}\vec{\sigma_{i}}.\vec{\sigma_{j}}\label{eq:A4}
\end{equation}
with $\vec{\sigma_{i}}$ the Pauli spin matrices of the $i^{th}$
quark and 
\begin{equation}
E_{ij}=8\alpha_{c}\frac{\mu_{i}(R)\mu_{j}(R)}{R^{3}}I_{ij}\label{eq:A5}
\end{equation}
where $\alpha_{c}$ is the color coupling constant. The magnetic moment
has the expression \begin{equation}
\mu_{i}(R)=\frac{R}{6}\frac{4\Omega(m_{i})+2Rm_{i}-3}{2\Omega(m_{i})\left(\Omega(m_{i})-1\right)+Rm_{i}}.\label{eq:A6}\end{equation}
and we point out that its quark mass dependence will produce a non
trivial flavor dependence of the coupling to the nuclear scalar field.
The expression for the overlap integral $I_{ij}$ is:\begin{eqnarray}
I_{ij} & = & 1+\frac{-3y_{i}y_{j}-4x_{i}x_{j}\sin^{2}x_{i}\sin^{2}x_{j}+x_{i}x_{j}K_{ij}}{2(x_{i}\sin^{2}x_{i}-\frac{3}{2}y_{i})(x_{j}\sin^{2}x_{j}-\frac{3}{2}y_{j})}\label{eq:A7}\\
K_{ij} & = & 2x_{i}Si(2x_{i})+2x_{j}Si(2x_{j})\\
&&-(x_{i}+x_{j})Si(2x_{i}+2x_{j})-(x_{i}-x_{j})Si(2x_{i}-2x_{j})\label{eq:A8}\\
y_{i} & = & x_{i}-\sin x_{i}\cos x_{i},\,\,\, x_{i}=\sqrt{\Omega(m_{i})^{2}-(Rm_{i})^{2}}\label{eq:A9}\end{eqnarray}
Using the spin-flavor wave functions of the baryons one can write:\begin{equation}
\Delta E_{M}=a_{00}E_{00}+a_{0s}E_{0s}+a_{ss}E_{ss}\label{eq:A10}\end{equation}
where the index 0 refers to the $u,d$ quarks ans $s$ to the strange
quark. The spin-flavor factors $a_{ii}$ are given in Table \ref{cap:hyperfine}.
\begin{table}
\begin{center}\begin{tabular}{|c|c|c|c|}
\hline 
&
$a_{00}$&
$a_{0s}$&
$a_{ss}$\tabularnewline
\hline
\hline 
N&
-3&
0&
0\tabularnewline
\hline 
$\Lambda$&
-3&
0&
0\tabularnewline
\hline 
$\Sigma$&
1&
-4&
0\tabularnewline
\hline 
$\Xi$&
0&
-4&
1\tabularnewline
\hline
\end{tabular}\end{center}

\caption{\label{cap:hyperfine}Spin-flavor factors for Eq.(\ref{eq:A10})}
\end{table}

To fix the free parameters $(B,Z_{0},m_{s},\alpha_{c})$ we take 
the free nucleon radius $R_{N}^{free}$ as the free parameter and
require that the nucleon and $\Delta(1232)$ have their physical
masses%
\footnote{The expression for the  energy of $\Delta(1232)$ is the same as for the nucleon except
that $a_{00}=3.$ %
}. Together with the stability equations this determines $(B,Z_{0},\alpha_{c})$.
We then choose $m_{s}$ so as to have a best fit to the $(\Lambda,\Sigma)$
masses. The resulting parameters are shown in Table \ref{FreeParameters}
for typical values of $R_{N}^{free}$.

\begin{table}
\begin{center}\begin{tabular}{|c|c|c|c|c|}
\hline 
$R_{N}^{free}({\rm\ fm})$&
$B({\rm\ fm}^{-4})$&
$Z_{0}$&
$\alpha_{c}$&
$m_{s}({\rm\ MeV})$\tabularnewline
\hline
\hline 
0.8&
0.554&
2.642&
0.448&
341\tabularnewline
\hline 
1.0&
0.284&
1.770&
0.560&
326\tabularnewline
\hline
\end{tabular}\end{center}

\caption{\label{FreeParameters}Values of the parameters $(B,Z_{0},m_{s},\alpha_{c})$
for typical values of $R_{N}^{free}$.}
\end{table}

\subsection{Effective mass}

We now consider the effect of the nuclear scalar field which is assumed
to be constant over the volume of the baryon. We recall that the variation 
of the field just produces the spin-orbit interaction, which
we can neglect when considering uniform matter. Let $\sigma$ be the
uniform value of the scalar field and $g_{\sigma}^{q}$ its coupling
to the $u,d$ quarks. We assume that the strange quark does not interact
with the scalar field. Therefore the coupling to the scalar field
amounts to the substitution 
\begin{equation}
m_{ud}\rightarrow m_{ud}-g_{\sigma}^{q}\sigma,\,\,\, m_{s}\rightarrow m_{s}
\label{eq:A11}
\end{equation}
in the free mass equation, Eq.~(\ref{eq:A1}), followed by application of
the stability
condition to determine the actual radius. The mass then becomes a
function of $\sigma$ and of the parameters, $(B,Z_{0},m_{s},\alpha_{c})$,  
which themselves depend only on $R_{N}^{free}$. So we have
\begin{equation}
M_{f}({\rm in\, medium})\equiv M_{f}(m_{ud}-g_{\sigma}^{q}\sigma,R_{N}^{free})
\label{eq:A12}
\end{equation}
Since we have two independent energy scales, $\sigma$ and $1/R_{N}^{free}$,
dimensional analysis is not very helpful and we simply make a fit
of $M_{f}(\sigma,R_{N}^{free})-M_{f}(0,R_{N}^{free})$ in powers of
$m_{ud}-g_{\sigma}^{q}\sigma$, with the coefficients also fitted as
polynomials in $R_{N}^{free}$. We then define the free $\sigma-N$
coupling as 
\begin{equation}
g_{\sigma}=\left.\frac{\partial M_{N}(\sigma,R_{N}^{free})}{\partial\sigma}
\right|_{\sigma=0}=-g_{\sigma}^{q}\left.
\frac{\partial M_{N}(\sigma,R_{N}^{free})}{\partial m}\right|_{m=0}
\label{eq:A13}
\end{equation}
which allows us to eliminate $g_{\sigma}^{q}$ in favor of $g_{\sigma}$.
We then get an expansion of the form
\begin{equation}
M_{f}(\sigma,R_{N}^{free})-M_{f}(0,R_{N}^{free})=
P_{f}^{(1)}(R_{N}^{free})g_{\sigma}\sigma+P_{f}^{(2)}(R_{N}^{free})(g_{\sigma}
\sigma)^{2}+....
\label{eq:A14}
\end{equation}
where, by construction
\begin{equation}
P_{N}^{(1)}(R_{N}^{free})=-1.
\label{eq:A15}
\end{equation}
If the mass were approximated by 
\begin{equation}
M_{f}=\frac{N_{u,d}\Omega(m_{ud})+N_{s}\Omega(m_{s})}{R_{N}^{free}}
\label{eq:A16}
\end{equation}
we would have
\begin{equation}
P_{\Lambda\Sigma}^{(1)}=-2/3,\,\,\, P_{\Xi}^{(1)}=-1/3,
\label{eq:A17}
\end{equation}
but this approximation is severely broken in the realistic case. The
order of the polynomials in $R$ has been set to 2 and the coefficients
have been  fitted over the range $[0.7{\rm\ fm}\rightarrow1.3{\rm\ fm}].$ 
We
have checked that an expansion truncated at order 
$\left(g_{\sigma}\sigma\right)^{2}$
was sufficient, even for scalar fields as large as 
$g_{\sigma}\sigma=600{\rm\ MeV}$.
This corresponds to densities of order $1{\rm\ fm}^{-3}$, 
which is sufficient
for our purposes. The expansion becomes, with everything expressed
in ${\rm\ fm}$:
\begin{eqnarray}
M_{N}(\sigma) & = & M_{N}-g_{\sigma}\sigma\nonumber \\
& &+\left[0.0022+0.1055R_{N}^{free}-
0.0178\left(R_{N}^{free}\right)^{2}\right]
\left(g_{\sigma}\sigma\right)^{2}
\label{eq:A18}\\
M_{\Delta}(\sigma) & = & M_{\Delta}-\left[0.9957-0.22737R_{N}^{free}+
0.01\left(R_{N}^{free}\right)^{2}\right]g_{\sigma}\sigma
\nonumber \\
 &  & +\left[0.0022+0.1235R_{N}^{free}-0.0415\left(R_{N}^{free}\right)^{2}
\right]\left(g_{\sigma}\sigma\right)^{2}
\label{eq:A19}\\
M_{\Lambda}(\sigma) & = & M_{\Lambda}-\left[0.6672+0.0462R_{N}^{free}-
0.0021\left(R_{N}^{free}\right)^{2}\right]g_{\sigma}\sigma
\nonumber \\
 &  & +\left[0.0016+0.0686R_{N}^{free}-0.0084\left(R_{N}^{free}\right)^{2}
\right]\left(g_{\sigma}\sigma\right)^{2}
\label{eq:A20}\\
M_{\Sigma}(\sigma) & = & M_{\Sigma}-\left[0.6706-0.0638R_{N}^{free}-
0.008\left(R_{N}^{free}\right)^{2}\right]g_{\sigma}\sigma
\nonumber \\
 &  & +\left[-0.0007+0.0786R_{N}^{free}-0.0181\left(R_{N}^{free}\right)^{2}
\right]\left(g_{\sigma}\sigma\right)^{2}
\label{eq:A21}\\
M_{\Xi}(\sigma) & = & M_{\Xi}-\left[0.3395+0.02822R_{N}^{free}-
0.0128\left(R_{N}^{free}\right)^{2}\right]g_{\sigma}\sigma
\nonumber \\
 &  & +\left[-0.0014+0.0416R_{N}^{free}-0.0061\left(R_{N}^{free}\right)^{2}
\right]\left(g_{\sigma}\sigma\right)^{2}
\label{eq:A22}
\end{eqnarray}

Note that in these calculations we use the physical masses for the free
particles. The influence of the bag model dependence then appears
only in the interaction piece, $M_{f}(\sigma)-M_{f}(0)$. It is convenient
to write the dynamical mass in the form
\begin{equation}
M_{f}(\sigma)=M_{f}-w_{f}^{\sigma}g_{\sigma}\sigma+\frac{d}{2}
\tilde{w}_{f}^{\sigma}\left(g_{\sigma}\sigma\right)^{2}
\label{eq:A23}
\end{equation}
where the scalar polarisability, $d$, and the dimensionless weights,  
$w_{f}^{\sigma},\tilde{w}_{f}^{\sigma}$, are deduced from 
Eqs.~(\ref{eq:A18}-\ref{eq:A22}).
For illustrative purposes we give their values in Table~\ref{cap:weights}
for our preferred value of the free radius, $R_{N}^{free}=0.8{\rm\ fm}$.

\begin{table}
\begin{center}\begin{tabular}{|c|c|c|c|c|}
\hline 
&
$N$&
$\Lambda$&
$\Sigma$&
$\Xi$\tabularnewline
\hline
\hline 
d$({\rm\ fm})$&
0.15&
0.15&
0.15&
0.15\tabularnewline
\hline 
$w_{f}^{\sigma}$&
1&
0.703&
0.614&
0.353\tabularnewline
\hline 
$\tilde{w}_{f}^{\sigma}$&
1&
0.68&
0.673&
0.371\tabularnewline
\hline
\end{tabular}\end{center}

\caption{\label{cap:weights}The weights for $R_{N}^{free}=0.8{\rm\ fm}$. }
\end{table}


%

\begin{thebibliography}{10}

\bibitem{lat00}
J.~M. Lattimer and M.~Prakash.
\newblock {\em Phys. Rep.}, 334:121, 2000.

\bibitem{hei00}
H.~Heiselberg and M.~Hjorth-Jensen.
\newblock {\em Phys. Rep.}, 328:328, 2000.

\bibitem{lat01}
J.~M. Lattimer and M.~Prakash.
\newblock {\em ApJ}, 550:426, 2001.

\bibitem{rik03}
J.~Rikovska Stone, J.~C. Miller, R.~Koncewicz, P.~D. Stevenson, and M.~R.
  Strayer.
\newblock {\em Phys. Rev.}, C68:034324, 2003.

\bibitem{lat05a}
J.~M. Lattimer and M.~Prakash.
\newblock {\em Phys. Rev. Lett.}, 94:111101, 2005.

\bibitem{lat05b}
J.~M. Lattimer and B.~F. Schultz.
\newblock {\em ApJ}, 629:979, 2005.

\bibitem{pod05}
Ph. Podsiadlowski, J.~D.~M. Dewi, P.~Lesaffre, J.~C. Miller, W.~G. Newton, and
  J.~R. Stone.
\newblock {\em Mon. Not. R. Astron. Soc.}, 361:1243, 2005.

\bibitem{kla06}
T.~Klahn et~al.
\newblock {\em PR}, C74:035802, 2006.

\bibitem{Weber:2004kj}
Fridolin Weber.
\newblock Strange quark matter and compact stars.
\newblock {\em Prog. Part. Nucl. Phys.}, 54:193--288, 2005, astro-ph/0407155.

\bibitem{Schaffner-Bielich:2004ch}
Jurgen Schaffner-Bielich.
\newblock Strange quark matter in stars: A general overview.
\newblock {\em J. Phys.}, G31:S651--S658, 2005, astro-ph/0412215.

\bibitem{Maruyama:2005yp}
Tomoyuki Maruyama, Takumi Muto, Toshitaka Tatsumi, Kazuo Tsushima, and
  Anthony~W. Thomas.
\newblock Kaon condensation and lambda nucleon loop in the relativistic
  mean-field approach.
\newblock {\em Nucl. Phys.}, A760:319--345, 2005, nucl-th/0502079.

\bibitem{Tatsumi:1995is}
Toshitaka Tatsumi.
\newblock Kaon condensation and neutron stars.
\newblock {\em Prog. Theor. Phys. Suppl.}, 120:111--134, 1995.

\bibitem{Lawley:2006ps}
S.~Lawley, W.~Bentz, and A.~W. Thomas.
\newblock Nucleons, nuclear matter and quark matter: A unified {NJL} approach.
\newblock {\em J. Phys.}, G32:667--680, 2006, nucl-th/0602014.

\bibitem{Lawley:2005ru}
S.~Lawley, W.~Bentz, and A.~W. Thomas.
\newblock The phases of isospin asymmetric matter in the two flavor {NJL}
  model.
\newblock {\em Phys. Lett.}, B632:495--500, 2006, nucl-th/0504020.

\bibitem{gui04}
P.~A.~M. Guichon and A.~W. Thomas.
\newblock {\em Phys. Rev. Lett.}, 93:132502, 2004.

\bibitem{gui06}
P.~A.~M. Guichon, H.~H. Matevosyan, N.~Sandulescu, and A.~W. Thomas.
\newblock {\em Nucl. Phys.}, A772:1, 2006.

\bibitem{gui88}
P.~A.~M. Guichon.
\newblock {\em Phys. Lett.}, B200:235, 1998.

\bibitem{Bissey:2005sk}
F.~Bissey et~al.
\newblock {\em Nucl. Phys. Proc. Suppl.}, 141:22--25, 2005, hep-lat/0501004.

\bibitem{gui96}
P.~A.~M. Guichon, K.~Saito, E.~N. Rodionov, and A.~W. Thomas.
\newblock {\em Nucl. Phys.}, A601:349, 1996.

\bibitem{Krein:1998vc}
G.~Krein, Anthony~W. Thomas, and K.~Tsushima.
\newblock {\em Nucl. Phys.}, A650:313--325, 1999, nucl-th/9810023.

\bibitem{Thomas:1982kv}
Anthony~W. Thomas.
\newblock Chiral symmetry and the bag model: A new starting point for nuclear
  physics.
\newblock {\em Adv. Nucl. Phys.}, 13:1--137, 1984.

\bibitem{chan06}
G.~Chanfray.
\newblock {\em private communication}.

\bibitem{gle00}
N.~K. Glendenning.
\newblock In {\em Compact Stars}. Springer, Berlin, Heidelberg, New York, 2000.

\bibitem{arn77}
W.~D. Arnett and R.~L. Bowers.
\newblock {\em Astrophys. J. Supplement}, 33:415, 1977.

\bibitem{pan71}
V.~R. Pandharipande.
\newblock {\em Nucl. Phys.}, A178:123, 1971.

\bibitem{bal99}
S.~Balberg, I.~Lichtenstadt, and G.~B. Cook.
\newblock {\em Astrophys. J. Supplement}, 121:515, 1999.

\bibitem{wir93}
R.~B. Wiringa.
\newblock {\em Rev. Mod. Phys.}, 65:231, 1993.

\bibitem{hof01}
F.~Hofmann, C.M. Keil, and H.~Lenske.
\newblock {\em Phys. Rev.}, C64:025804, 2001.

\bibitem{iid98}
K.~Iida and K.~Sato.
\newblock {\em Phys. Rev.}, C58:2538, 1998.

\bibitem{bur02}
G.~F. Burgio, P.~K.~Sahu M.~Baldo, and H.-J. Schulze.
\newblock {\em Phys. Rev.}, C66:025802, 2002.

\bibitem{men03}
D.~P. Menezes and C.~Providencia.
\newblock {\em Phys. Rev.}, C68:035804, 2003.

\bibitem{bay71a}
G.~Baym, H.~Bethe, and C.~Pethick.
\newblock {\em Nucl. Phys.}, A175:225, 1971.

\bibitem{bay71b}
G.~Baym, C.~Pethick, and P.~Sunderland.
\newblock {\em ApJ}, 170:299, 1971.

\bibitem{nic05}
D.~J. Nice, E.~M. Splaver, I.~H. Stairs, O.~L\"ohmer, A.~Jessner, M.~Kramer,
  and J.~M. Cordes.
\newblock {\em ApJ}, 634:1242, 2005.

\bibitem{akm98}
A.~Akmal, V.~R. Pandharipande, and D.~G. Ravenhall.
\newblock {\em Phys. Rev.}, C58:1804, 1998.

\bibitem{cha97}
E.~Chabanat, P.~Bonche, P.~Hansel, J.~Meyer, and R.~Schaeffer.
\newblock {\em Nucl. Phys.}, A627:710, 1997.

\bibitem{tol34}
R.~C. Tolman.
\newblock {\em Proc. Natl. Acad. Sci. USA}, 20:3, 1934.

\bibitem{opp39}
J.~R. Oppenheimer and G.~M. Volkov.
\newblock {\em Phys. Rev.}, 55:374, 1939.

\bibitem{sch87}
R.~Schaffer, Y.~Declais, and S.~Julian.
\newblock {\em Nature}, 300:142, 1987.

\bibitem{hes06}
J.~W.~T. Hessels, S.~C. Ransom, I.~H. Stairs, P.~C.~C. Freire, V.~M. Kaspi, and
  F.~Camilo.
\newblock {\em Science}, 311:1901, 2006.

\bibitem{han89}
P.~Haensel and J.~L. Zdunik.
\newblock {\em Nature (London)}, 340:617, 1989.

\bibitem{han94}
P.~Haensel, M.~Salgado, and S.~Bonazzola.
\newblock {\em Astron. Astrophys.}, 296:745, 1994.

\bibitem{har67}
J.~B. Hartle.
\newblock {\em ApJ}, 150:1005, 1967.

\bibitem{rik02}
J.~Rikovska Stone, P.~D. Stevenson, J.~C. Miller, and M.~R. Strayer.
\newblock {\em Phys. Rev.}, C65:064213, 2002.

\bibitem{bet74}
H.~A. Bethe and M.~B. Johnson.
\newblock {\em Nucl. Phys.}, A230:1, 1974.

\bibitem{pag06}
D.~Page, U.~Geppert, and F.~Weber.
\newblock {\em in print in Nucl. Phys. A}, 2006.

\bibitem{kit06}
F.~S. Kitaura, H.-Th. Janka, and W.~Hillebrant.
\newblock 2006.

\bibitem{sta06}
H.~Stairs, S.~E. Thorsett, R.~J. Dewey, M.~Kramer, and C.~A. McPhee.
\newblock 2006.

\bibitem{pag04}
D.~Page, J.~M. Lattimer, M.~Prakash, and A.~W. Steiner.
\newblock {\em Astrophys. J. Suppl.}, 155:623, 2004.

\bibitem{jon01}
P.~B. Jones.
\newblock {\em Phys. Rev.}, D64:084003, 2001.

\bibitem{men04}
D.~P. Menezes and C.~Providencia.
\newblock {\em Phys. Rev.}, C69:045801, 2004.

\bibitem{DeGrand:1975cf}
Thomas~A. DeGrand, R.~L. Jaffe, K.~Johnson, and J.~E. Kiskis.
\newblock {\em Phys. Rev.}, D12:2060, 1975.


\end{thebibliography}

\end{document}